\documentclass[%
 reprint,
superscriptaddress,
 amsmath,amssymb,
 aps,
]{revtex4-2}

\usepackage{graphicx}
\usepackage{dcolumn}
\usepackage{bm}
\usepackage{gensymb}

\begin{document}
\preprint{AAPM/123-QED}

\title[Binary Volume Acoustic Holograms]{Binary Volume Acoustic Holograms}

\author{Michael D. Brown}
 \email{Michael.brown.13@ucl.ac.uk}
 \affiliation{ 
Department of Medical Physics and Biomedical Engineering, University College London, Gower St, London, WC1E 6BT
}%
\affiliation{Department of Neuroscience,\ Erasmus\ MC,\ Rotterdam,\ Netherlands}
\author{Ben T. Cox}
\affiliation{ 
Department of Medical Physics and Biomedical Engineering, University College London, Gower St, London, WC1E 6BT
}%
\author{Bradley E. Treeby}
\affiliation{ 
Department of Medical Physics and Biomedical Engineering, University College London, Gower St, London, WC1E 6BT
}%

\date{\today}

\begin{abstract}
In recent years high-resolution 3D printing has enabled a diverse range of new, low-cost, methods for ultrasonic wave-front shaping. Acoustic holograms, particularly, allow for the generation of arbitrary, diffraction limited, acoustic fields at MHz frequencies from single element transducers. These are phase plates that function as direct acoustic analogues to thin optical holograms. In this work it is shown that, by using multiple polymer 3D printing, acoustic analogues to `thick' or volume optical holograms can also be generated. First, an analytic approach for designing a volume hologram that diffracts a set of input fields onto a desired set of output fields is briefly summarised. Next, a greedy optimisation approach based on random downhill binary search able to account for the constraints imposed by the chosen fabrication method is introduced. Finally, an experimental test-case designed to diffract the field generated by a 2.54 cm, planar, PZT transducer onto 8 distinct patterns dependent on the direction of the incident field is used to validate the approach and the design method. Field scans of the 8 target fields demonstrate that acoustic analogues of optical volume holograms can be generated using multi-polymer printing and that these allow the multiplexing of distinct fields onto different incident field directions. 
\end{abstract}

\maketitle

The ability to precisely shape and focus acoustic fields is essential to a range of applications in biomedical ultrasound\cite{Melde2018,Baudoin2020,Kruizinga2017,Maimbourg2018}. For transcranial ultrasound neuromodulation, for example, such control is required to precisely compensate for the distortions introduced by propagation through the skull\cite{Kyriakou2014}. In most existing systems these corrections are achieved via arrays of independent sources, however, increasingly, there is interest in the potential of passive acoustic metamaterials\cite{Memoli2017} and lenses\cite{Maimbourg2018,Jimenez-Gambin2019} as a cheaper, simpler, alternative for transcranial focusing and other biomedical applications. Promising among these has been the development acoustic holograms\cite{Lalonde1991,Melde2018}. Holograms are phase plates, often 3D printed, that can map the continuous wave output of a single element transducer onto a pre-computed phase pattern which is designed to diffract to form a desired field. 

Acoustic holograms are cheap to fabricate and offer higher spatial fidelity over the transmitted field than traditional arrays, however, they are static. Each hologram generates a fixed distribution, as such, their flexibility is limited. To circumvent this, two methods have been introduced. Spatial steering and scaling of a fixed field by modifying the driving frequency or through the use of a coarse array has been reported in several works\cite{Memoli2019,Cox2019,Lalonde1995}. However, the range available to frequency-based steering is limited by the increasing coupling of energy into different diffraction orders\cite{Brown2020}, while steering using an array requires more complex driving electronics and is subject to distortions from grating lobes if insufficiently sampled and is redundant for fully-sampled arrays. Alternatively, the frequency dependence of the phase delay introduced by a hologram has been exploited to multiplex distinct patterns onto different driving frequencies\cite{Brown2017}. However, this approach is limited by cross-talk between target patterns and the limited transducer bandwidth. Recent works have demonstrated dynamic equivalents to these holographic lenses in the form of re-write-able bubble masks generated via electrolysis or optically induced electro-chemical interactions\cite{Ma2020,Ma2022}. However, these approaches have higher cost and technical complexity and, at present, have only demonstrated the capacity for binary amplitude modulation of an incident field limiting the efficiency with which energy can be coupled into the desired field to 10$\%$\cite{Moreno1997}.

Each of these prior works have investigated 'thin' holographic lenses (Fig.\ 1 (a)), with a thickness $d$ on the order of the acoustic wavelength $\lambda$, to modulate the transmitted field. Both the design and acoustical behaviour of these lenses is directly analogous to thin optical holograms (e.g., as generated using a spatial light modulator). In this work, it is demonstrated that by exploiting multiple polymer 3D printing it is also possible to generate binary volume acoustic holograms. These are direct acoustic analogues to volume optical holograms also known as thick or Bragg holograms (Fig.\ 1(b)). Rather than thin phase elements, volume holograms are extended structures ($d >> \lambda$), in which the refractive index or absorption modulates periodically about some bulk value. This extended thickness results in holograms that are significantly more selective to frequency/direction of the incident field as only waves that match the Bragg criteria for the hologram diffract\cite{Russell1981}. As a result, the ability to multiplex distinct patterns within a single structure is significantly greater (e.g., for optical information storage\cite{VanHeerden1963}). This enhanced ability to multiplex in a single fixed lens could improve the flexibility of acoustic holograms for biomedical applications.

\begin{figure}[t!]
    \centering
    \includegraphics[width=\columnwidth]{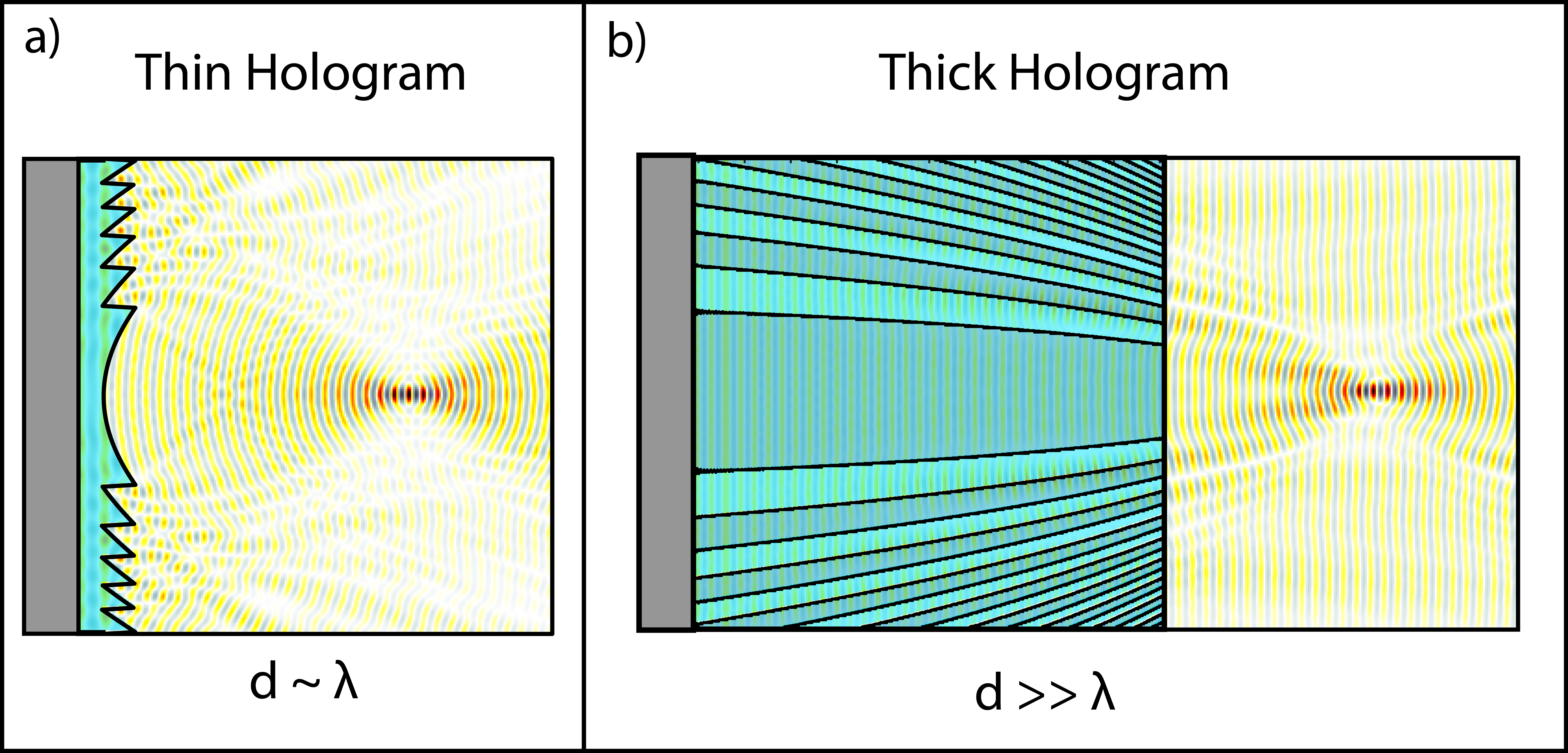}
    \caption{(a) Visualisation of a thin acoustic hologram which shapes the field acting as a thin phase screen. (b) Visualisation of a volume acoustic hologram which shapes the field through scattering from sound speed variations.}
    \label{fig_traces}
\end{figure}

\section{Volume Hologram Design}
\subsection{Problem statement}
The general challenge in designing a volume acoustic hologran can be framed as finding a 3D sound speed distribution $c(\mathbf{r})$ that maps a set of $n$ input fields $p_i^n(\mathbf{r})$ that vary spatially or spectrally onto a set of $n$ predefined output fields $p_o^n(\mathbf{r})$, where $\mathbf{r}$ is the spatial coordinate $(x,y,z)$. This could be treated analogously to Ultrasound Tomography (UST)\cite{Perez-Liva2017} by using full-waveform inversion to find a set of acoustic properties that minimise the residual between a desired field and the output of a numerical wave equation solver. However, incorporating design constraints such that the values converge to a stable solution that can be physically realised would pose a significant challenge. As such, for this work, a single-scattering (Born) approximation is employed thereby constraining the problem to searching for a sound speed distribution $c(\mathbf{r})$ with an average bulk value $c_m$ and a weak perturbation $c_h(\mathbf{r})$ around this bulk value. That is 
\begin{equation}
    c(\mathbf{r}) = c_m + c_h(\mathbf{r}).
\end{equation}
The goal then is to find this perturbation $c_h(\mathbf{r})$ that maps between our desired input-output fields. This re-framing allows for the design problem to be significantly simplified.

\subsection{Lossless Volume Hologram Design}
The behaviour of volume hologram gratings has been widely studied in optics, which can be used to inform the design. The earliest rigorous theoretical treatment is coupled wave theory developed by H. Kogelnik in the 1960s\cite{Kogelnik1969} which describes the Bragg diffraction of waves from an infinite slab with finite thickness containing periodic modulations in the refractive index or absorption.  

For the simplest case of a transmission hologram mapping an incident plane-wave $p_i(\mathbf{r}) = e^{i\mathbf{k_i\cdot r}}$ with wavevector $\mathbf{k_i}$ onto an output plane-wave $p_o(\mathbf{r}) = e^{i\mathbf{k_o\cdot r}}$ with wavevector $\mathbf{k_o}$, a medium $c_h(\mathbf{r})$ coupling these two fields can be shown to be
\begin{equation}
    c_h(\mathbf{r}) = \Re\{e^{i\mathbf{k_{i}\cdot r}}e^{-i\mathbf{k_{o}\cdot r}}\Delta c\}.
\end{equation}

\noindent Here $\Delta c$ is the scaling of the perturbation. Equation (2) defines a new grating with a wavevector $\mathbf{k_g}$ given by the difference of the two wave-vectors
\begin{equation}
    \mathbf{k_g} = \mathbf{k_i}-\mathbf{k_o}.
\end{equation}

\begin{figure}
    \centering
    \includegraphics[width=\columnwidth]{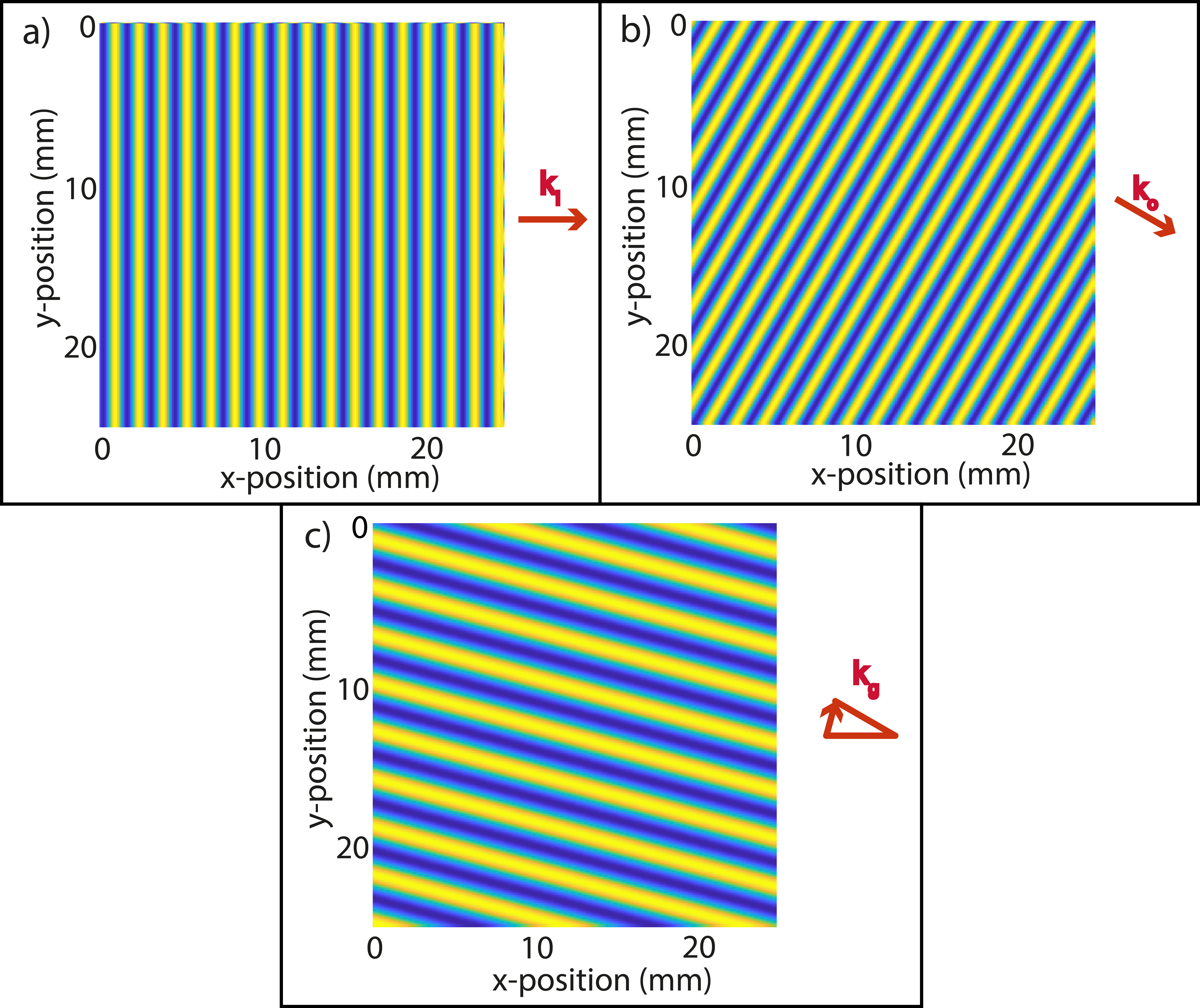}
    \caption{(a) Real component of incident plane wave with $\theta_i = 0\degree$, frequency 2 MHz. (b) Real component of output plane wave with $\theta_i = 30\degree$, frequency 2 MHz. (c) Sound speed grating coupling the input (a) and output (b) fields.}
    \label{fig_traces}
\end{figure}
\noindent This case is illustrated by Fig. 2, which shows the real component of an incident field $p_i$ (a), output field $p_o$ (b), and hologram medium $c_h$ (c) for a case where $\mathbf{k_i}$ is parallel to $x$ (i.e., $\theta_i = 0\degree$) and $\mathbf{k_o}$ has an angle $\theta_o = 30\degree$ relative to $x$. Here, the domain size is 25x25 mm, the frequency $f$ is 2 MHz, and the bulk sound speed $c_m$ is 1484 m$\cdot$s$^{-1}$. 

Assuming the grating starts in the plane $x=0$, and the incident field is the input wave $p_i(\mathbf{r})$, the fraction of energy $\eta$ in the output wave $p_o(\mathbf{r})$ as a function of depth $x$ can be calculated, from coupled wave theory, using\cite{BrothertonRatcliffei2013}

\begin{equation}
    \eta (x) = \sin^2(\frac{\Delta c}{2c_m}kx(\sec(\theta_i)\sec(\theta_o))^{\frac{1}{2}}). 
\end{equation}

\noindent Here $k = |\mathbf{k_i}|$ is the wavenumber, $\theta_i$ is the angle of the incident wave relative to $x$, $\theta_o$ is the angle of the output wave relative to $x$, and $sec$ is the secant function. This assumes that the perturbation $\Delta c < c_m$ (so there is no coupling into higher-diffraction orders) and only a single pair of coupled waves. More rigorous models able to exactly account for such effects (e.g., coupling into higher-diffraction orders or multiple super-imposed gratings) have been reported\cite{Moharam1981,Russell1981}. Equation 4 demonstrates that the fraction of energy coupled into the output field by a volume hologram grating is determined by the angles of the two plane waves, the magnitude of the sound speed perturbation, the frequency, and the thickness of the hologram. More generally, with depth, the hologram $c_h$ first couples energy from the incident wave $p_i$ into the output wave $p_o$ this continues until the incident wave is fully depleted (i.e., all the energy is contained in $p_o$) after which the grating couples energy back into $p_i$. This process repeats continually as a function of depth within the hologram. 

This analytic model was compared to a simulation carried out in 2D using the k-Wave toolbox, a pseudo-spectral model for time-domain wave simulations.\cite{Treeby2010} The contrast magnitude $\Delta c$ was set to 40 m$\cdot$s$^{-1}$, the grid spacing was 200 $\mu$m, the grid dimensions were 100x100 mm, and the temporal spacing was 27 ns. The fields $p_i$, $p_o$, and grating $c_h$ were as shown in Fig.\ 2. The input field $p_i$ was inserted as a continuous wave, planar, additive source at 2 MHz along $x=0$ with a width of 40 mm. This was spatially apodised using a Hanning window. The magnitude of the resulting steady state field at 2 MHz is shown in Fig. 3(a). The coupling between $p_i$ and $p_o$ as a function of depth $x$ can clearly be seen. The fraction of energy in $p_i$ and $p_o$ was evaluated using a 1D spatial Fourier transform for each depth. The normalised variation of energy $p_o$ compared to the prediction of coupled wave theory (Eq.\ (4)) is shown in Fig.\ 3(b), the two agree near exactly, showing that the underlying approximations of coupled wave theory are valid for this $\Delta c$. 

\begin{figure}
    \centering
    \includegraphics[width=\columnwidth]{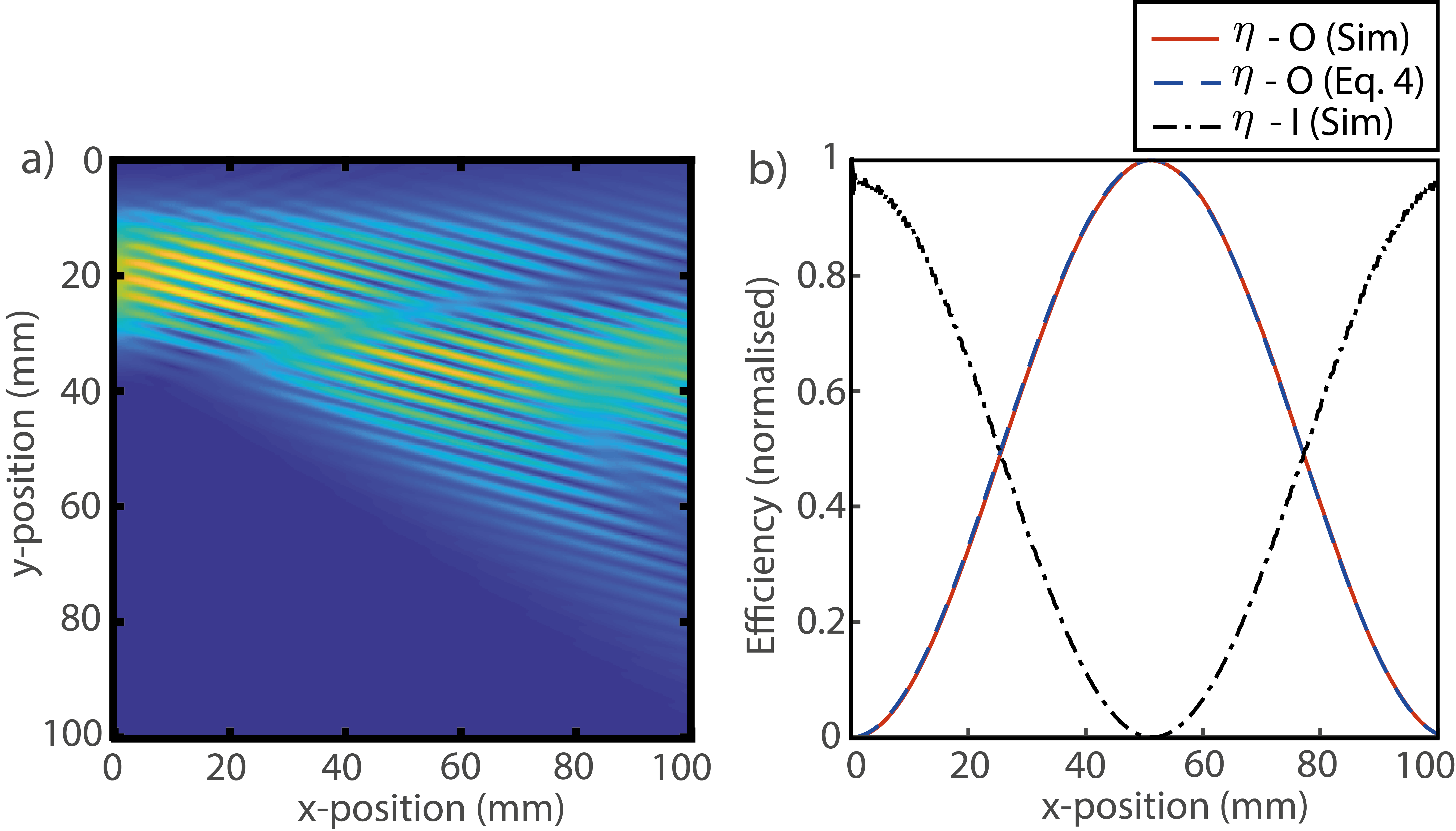}
    \caption{(a) Magnitude of steady state field at 2 MHz for simulated propagation of normally incident plane-wave inside extension of grating shown in (Fig. 2(c)). (b) Energy contained in $I$ and $O$ as a function of depth within grating along with theoretical prediction. Demonstrating that, for these example parameters, assumptions underlying coupled wave theory are valid.}
    \label{fig_traces}
\end{figure}

This approach for designing a volume hologram grating can be generalised to arbitrary combinations of input and output fields. Assuming the fields within the volume hologram are represented by their angular spectra $\hat{p}_i(k_x,k_y,k_z)$ and $\hat{p}_o(k_x,k_y,k_z)$ then a hologram coupling the two fields $c_h(x,y,z)$ can be calculated as a weighted sum of gratings (Eq. (2)) using

\begin{equation}
\begin{aligned}
           c_h(x,y,z) = \Re\Bigg\{\idotsint \\
           \hat{p}_i(k_{ix},k_{iy},k_{iz})e^{i(k_{ix}x+k_{iy}y+k_{iz}z))}\\
           \hat{p}_o(k_{ox},k_{oy},k_{oz})e^{-i(k_{ox}x+k_{oy}y+k_{oz}z))}\\
           dk_{ix}dk_{iy}dk_{iz}dk_{ox}dk_{oy}dk_{oz}.\Bigg\}
           \end{aligned}
\end{equation}
Equation (5) is the Fourier transform of the convolution of the two angular spectra so simplifies to a multiplication of the fields in the spatial domain. We can adopt this approach for low-contrast gratings owing to linearity under the Born approximation. 

\begin{figure*}
    \centering
    \includegraphics[width=1.6\columnwidth]{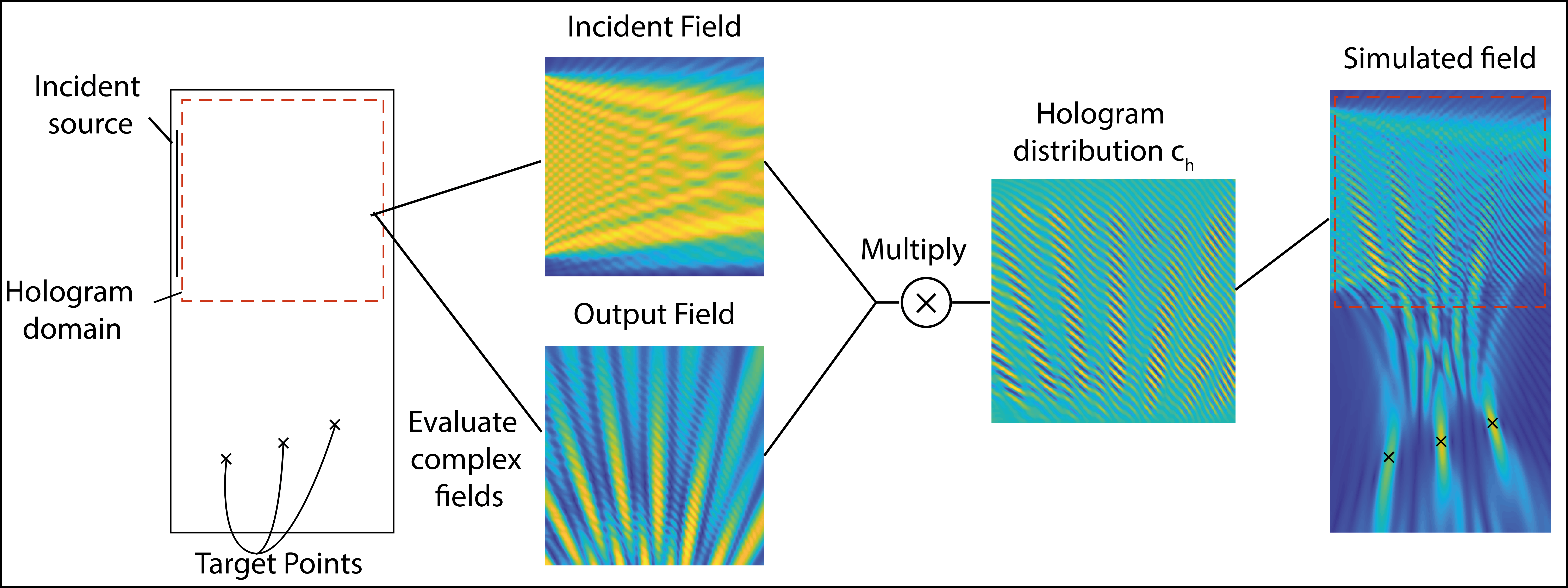}
    \caption{Illustration of approach to calculate volume hologram in non-absorbing, continuous, medium. First, the target field, input field and hologram domain are defined. Next, the incident and target output fields within the hologram domain are calculated, these are then multiplied to calculate $c_h$. Upon imposing $c_h$ in the simulation domain the hologram scatters the incident field onto the target field.}
    \label{fig_traces}
\end{figure*}

The calculation of a hologram with this approach for an arbitrary case is illustrated in Fig.\ 4. Here, the hologram is 2D with dimensions of 3x3 cm, the incident field is a planar source with a width of 2.5 cm, the target field is a set of 3 focal-points, outside the hologram, at differing lateral positions and depths. The design frequency is 2 MHz and $c_m$ is 1484 m$\cdot$s$^{-1}$. First, the incident and target fields at 2 MHz are calculated within the hologram boundaries. Next, these are spatially multiplied and the real component of the resulting distribution taken to give the target sound speed distribution $c_h$. Finally, upon imposing this sound speed distribution within the hologram domain the incident plane-wave is mapped onto the desired target distribution. 

A key properties of volume holograms is their capacity to multiplex distinct output fields in a single volume. Two types are typically considered; spatial (or angular) multiplexing where distinct outputs are encoded onto different incident fields (or fields incident from different angles)\cite{Heanue1994} and frequency multiplexing where they are encoded onto different driving frequencies\cite{Rosen1993}. Practically, using a single-scattering assumption, design of a hologram $c_h$ multiplexing between a distinct set of $N$ input $p^n_i(\mathbf{r})$ and output $p^n_o(\mathbf{r})$ fields can be achieved via a summation of the product of the individual fields. That is 
\begin{equation}
\begin{aligned}
           c_h(\mathbf{r}) = \Re\Bigg\{\sum_{n=1}^{N}            p^n_i(\mathbf{r})p^n_o(\mathbf{r})\Delta c\Bigg\}.
           \end{aligned}
\end{equation}

\noindent This, however, creates two challenges. First, the diffraction efficiency of each individual field drops by approximately $\sqrt{N}$\cite{Shishova2020}. Second, too narrow an angular or wavelength separation results in cross-talk between the different fields. From coupled wave theory, for the case of coupling two planar waves, the angular and wavelength selectivity are related to the dimensions of the grating and the angles of the incident and output waves via the following expressions\cite{Burr1996}
\begin{equation}
\begin{aligned}
           \Delta\theta_i = \frac{\lambda}{L}\frac{\cos{(\theta_o)}}{\sin{(\theta_i+\theta_o)}},
           \end{aligned}
\end{equation}
 \begin{equation}
\begin{aligned}
           \frac{\Delta\lambda}{\lambda} = \frac{\lambda}{L}\frac{\cos{(\theta_o)}}{1-\cos{(\theta_i+\theta_o)}}.
           \end{aligned}
\end{equation}
Here $\lambda = 2\pi/k$, and L is the hologram thickness. This, in principle, gives a high degree of selectivity. For example, for L = 3 cm, $c_m$ = 1484 m$\cdot$s$^{-1}$, $f$ = 2 MHz, $\theta_o$ = 30$\degree$, and $\theta_i$ = 0$\degree$. The wavelength selectivity $\Delta\lambda$ is 320 kHz while the angular selectivity $\Delta \theta_i$ is 2.45$\degree$. However, it should be noted that these equations are for a transmission geometry, other geometries (e.g., reflection or 90$\degree$) have differing behaviour to change in incidence and wavelength.\cite{Kogelnik1969} For more complex output fields the prediction of cross-talk is more involved. Requiring analysis of the Ewald spheres of the target fields\cite{Gerke2010}.

\section{Fabrication Requirements} 
The approach outlined in Sec.\ II can be used to calculate a volume hologram for diffracting between arbitrary input-output fields in an ideal case where the sound speed can be continuously varied and there is no loss. However, this isn't possible, at present, to physically realise. Fabrication of an acoustic volume hologram requires the ability to, controllably, spatially vary the mechanical properties of a medium on a length-scale on the order of the acoustic wavelength. At 2 MHz, for different plastics, this length-scale varies between ~0.5-1.3 mm while the mechanical contrast magnitude $\Delta c$ required varies with geometry. From Eq.\ 4 for a 3 cm hologram, with $\theta_i = 0\degree$, $\theta_o = 30\degree$, f = 2MHz, with $c_m$ = 2495 m$\cdot$s$^{-1}$ the contrast ratio $\frac{\Delta c}{c_m}$ needs to be approximately $2\%$ or $\Delta c$ = 50 ms$^{-1}$ for complete conversion of an incident wave (i.e., 100$\%$ diffraction efficiency).  

For this work multi-polymer polyjet 3D printing on a Stratasys J835 (Stratasys, Edina, MN, US) is used for fabrication. This allows for the printing of wide-range of materials in a single component with a spatial resolution of $<$200 $\mu$m. Specifically, the J835 prints `primary' materials Agilus30 (rubber-like) and veroClear (Poly(methyl methacrylate)-like (PMMA)) and synthesises a broad range of `derived' materials with mechanical properties varying between these two by depositing pre-determined mixtures of the two primary materials.\cite{Vaezi2013} In a recent work\cite{Bakaric2021}, the bulk acoustical properties of both of these primary materials as well as several of the derived materials (FLXA9960, FLXA9995, RGDA8625, and RGDA8630) were characterised, finding the group velocity between 1-3.5 MHz to vary between 2019-2496 m$\cdot$s$^{-1}$. However, these values weren't distributed continuously over this range instead being clustered around 2100 m$\cdot$s$^{-1}$ (FLXA9960, FLXA9995, and Agilus30) and 2450 m$\cdot$s$^{-1}$ (RGDA8625, RGDA8630, and VeroClear). In principle, two materials, RGDA8630 (2447 m$\cdot$s$^{-1}$) and veroClear (2475 m$\cdot$s$^{-1}$), have sufficient contrast to generate a volume hologram ($\Delta c = 14$ m$\cdot$s$^{-1}$). Practically, however, such a hologram was found to be extremely weakly diffracting. As such to approximate a desired contrast new 'mixed' materials were instead synthesised using random mixtures of Agilus30 ($c$ = 2035 m$\cdot$s$^{-1}$) and veroClear (2475 m$\cdot$s$^{-1}$) on a 200 $\mu m$ length scale. Setting the volume fraction of each material based on the empirically determined desired sound speed contrast (see Sec.\ V D).  

While multi-polymer printing, in principle, offers sufficient fidelity and contrast its use introduces two additional challenges. First, the losses due to attenuation within the photopolymers at 2 MHz range from 6.3-18 dB/cm.\cite{Bakaric2021} Second, the sound speed distributions that can be fabricated are quantised. For simplicity, in this work, the design of holograms comprised of just 2 materials or binary distributions are considered.  These two challenges prohibit the use of a simple lossless model (Sec.\ II) to design the hologram. Figure 5 shows the effect of both binarisation and attenuation on the hologram profile simulated in Fig.\ 4. The pressure generated over the target points drops and significant variability in the pressure generated over the 3 focal points is introduced. This variability in the pressure arises principally from attenuation which `apodises' the grating such that scattering from structures closer to the source are stronger. Simple thresholding of a continuous target distribution also results in over-weighting of edges and corners in a target pattern.\cite{Tsang2011} To account for both attenuation and binarisation a greedy optimisation approach was developed for the design of binary volumetric acoustic holograms in absorbing media. This is an adaptation of an algorithm introduced in several previous works\cite{Brown2014,Brown2017}. 

\begin{figure}
    \centering
    \includegraphics[width=\columnwidth]{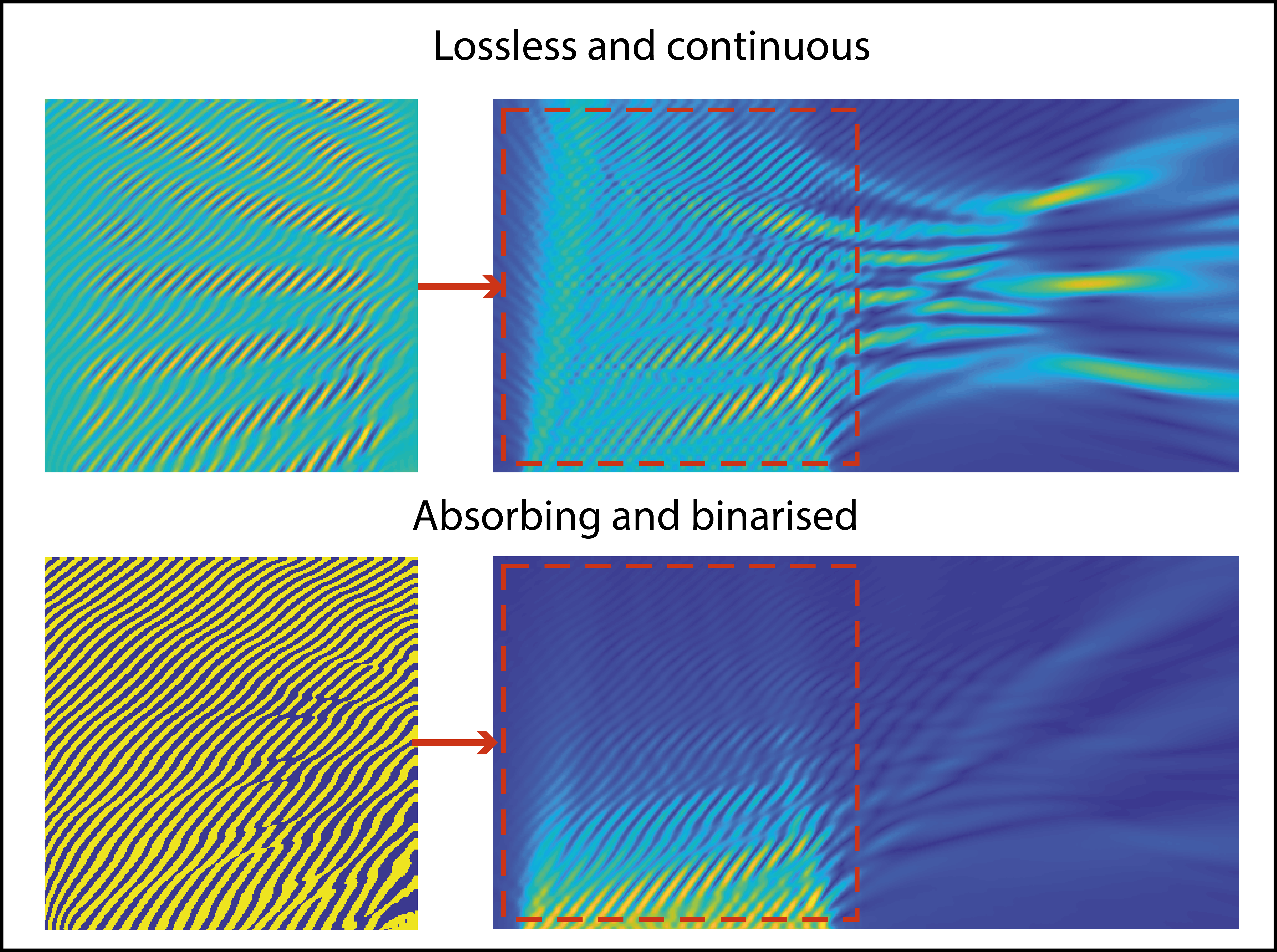}
    \caption{Illustration of the impact of absorption and binarisation on acoustic field generated by thick hologram. Attenuation `apodises' the grating such that scattering from structures close to the source are stronger. Binarisation over-weights edges and corners.}
    \label{fig_traces}
\end{figure}

\section{Greedy design algorithm}
The inverse problem now is to find a binary spatial distribution $h(\mathbf{r})$ that maps a set of continuous wave input fields described within the hologram as $p_i^n(\mathbf{r})$ onto a set of output target pressure distributions $p_o^n(\mathbf{r})$. These input fields can be defined either for a single frequency $f$ or for a set of distinct frequencies $f_n$, However, in this work only a single input frequency will be considered. 

The hologram is first discretised. For example, a cubic hologram $h(\mathbf{r})$ occupying an L$\times$L$\times$L region is represented as a set of evenly spaced points $\mathbf{r}_k = \{(x_k,y_k,z_k), k = 1,...,K\}$ with maximum spacing $\Delta < \frac{\lambda}{2}$ in each dimension to meet the sampling requirements. Here, $\lambda$ is the wavelength in the hologram medium, and the subscript $k$ is the coordinate index. For definition of arbitrary geometries the set of points, from a regular grid, are restricted to those that fall within the desired volume. The target distributions $p_o^n$ are also constrained to a single target plane $z_d$ external to the hologram (i.e., $p_o^n = p^n_o(x,y,z_d)$). Additionally, rather than continuous distributions, a set of target points $\mathbf{r}^n_m =\{(x_m,y_m,z_d), m = 1,...,M\}_n$ are instead defined for each output field.   

To solve this inverse problem it's necessary to have a forward model linking each input field $p_i^n(\mathbf{r})$ and hologram state $h(\mathbf{r})$ to the pressure $p_m^n$ generated at each target point $\mathbf{r}_m^n$. This is modelled as
\begin{equation}
    p_m^n = \sum_{k=1}^{K}G^n_m(\mathbf{r}_k)p^n_i(\mathbf{r}_k)h(\mathbf{r}_k).
\end{equation}

\noindent Here, $G^n_m(\mathbf{r}_k)$ is the Green's function between the $m^{th}$ target point of the $n^{th}$ target distribution and the position $\mathbf{r}_k$ inside the hologram, $p^n_i(\mathbf{r}_k)$ is the value of the $n^{th}$ input field at $\mathbf{r}_k$ inside the hologram, and $h(\mathbf{r}_k)$ is the hologram state at the position $\mathbf{r}_k$. 

Both $G^n_m(\mathbf{r}_k)$ and $p_i^n(\mathbf{r}_k)$ are pre-computed. We consider volume holograms that can have a complex geometry and that are designed to generate target distributions in an external medium (water) which has acoustic properties that differ significantly from the 3D printed photopolymers. As such it's necessary to model diffraction, reflection, and attenuation for waves coupling into and out of the hologram when evaluating $G^n_m(\mathbf{r}_k)$ and $p_i^n(\mathbf{r}_k)$. This is accomplished using the k-Wave toolbox, inserting $h(\mathbf{r})$ into the simulation domain as a region with the acoustic properties matching those of the printable photopolymers. However, fully modelling $h(\mathbf{r})$ would require updating $G^n_m(\mathbf{r}_k)$ and $p_i^n(\mathbf{r}_k)$ each time the hologram state is changed. For simplicity we instead approximate the hologram medium as homogeneous with a bulk sound speed, density, and attenuation $(c_m,\rho_m,\alpha_m)$ defined as the average of the two materials used to fabricate the hologram (i.e., the Greens functions are unaffected by the hologram state $h(\mathbf{r})$). This forward model makes a number of further simplifying assumptions namely:
\begin{enumerate}
  \item The hologram is single scattering
  \item The incident fields can be neglected at the target points (i.e., $p_i^n(\mathbf{r}^n_m) = 0$).
  \item The hologram is a fluid.
  \item There is no depletion of the incident field by the hologram.
  \item There is no density contrast between the materials.
\end{enumerate}
It is demonstrated experimentally in Sec.\ V that these assumptions are valid for certain sound speed contrasts and problem geometries.

The optimisation then searches for a binary volume hologram that maximises the pressure generated over the set of target points for each output field while minimising the variation in pressure by maximising the following cost function
\begin{equation}
    C = E(|p^n_m|) - \beta(std|p^n_m|).
\end{equation}
Here, $p^n_m$ is the pressure at each target point $\mathbf{r}^n_m$, $\alpha$ is a weighting parameter. The first term $E(|p_m^n|)$ is the average pressure over the set of target points for each input field. The second term is the standard deviation of the same parameter. This cost function aims to maximises the energy over each target distribution whiles minimising the variation of pressure. For this work the value of $\beta$ was empirically determined and set to 0.7.

The optimisation was solved using random downhill binary search. The hologram was initialised in a fully off state $h^0(\mathbf{r})$. From the state $h^{l}(\mathbf{r})$, the optimisation proceeds by selecting a voxel $\mathbf{r}_k$ at random (from those not yet evaluated), this voxel is assigned to the state that maximises the cost function (Eq.\ 7). This process then repeats until all voxels have been tested resulting in an updated hologram $h^{l+1}(\mathbf{r})$. This then repeats until the number of changes in an iteration falls below 0.5$\%$ of the number of voxels in the hologram or 5 iterations are completed. 

\section{Experimental Validation}
\subsection{Introduction}
A test case was fabricated to demonstrate that multi-polymer printing can be used to realise acoustic analogues of volume holograms and to validate the greedy design algorithm. This was designed to generate the numerals from `1' to `8' for incident fields from 8 different directions. 

\subsection{Geometry}
\begin{figure}
    \centering
    \includegraphics[width=\columnwidth]{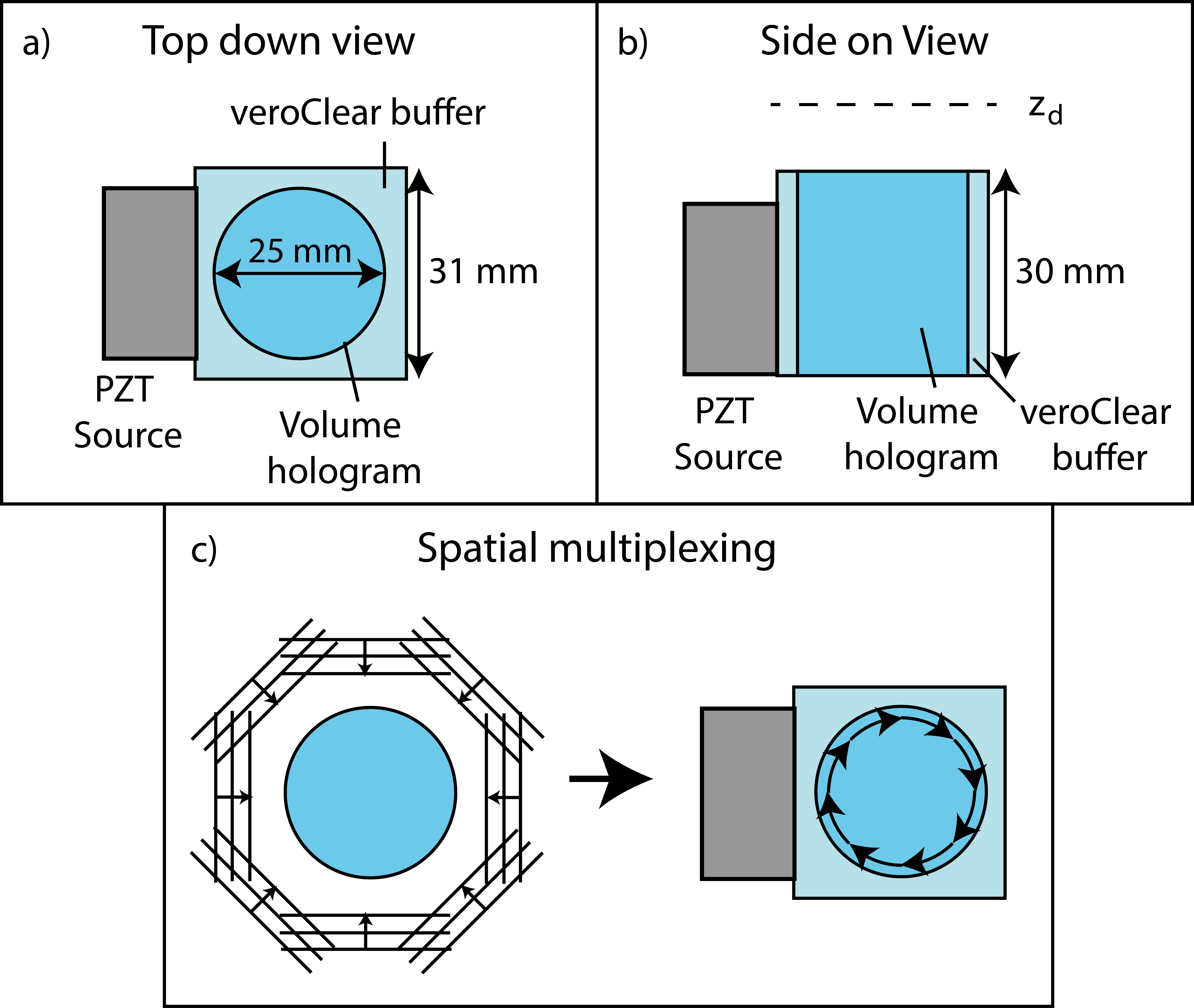}
    \caption{(a) Top-down view of geometry of experimental volume hologram. The test-case is a cylindrical hologram of diameter 25 mm embedded in a 31x31x30 cm cuboid buffer. (b) Side-on view of geometry of experimental volume hologram. (c) Illustration of approach used to spatially multiplex 8 patterns inside volume hologram geometry. The cylindrical hologram is able to rotate inside the buffer to change the direction of the incident field from the fixed transducer relative to the hologram.}
    \label{schem}
\end{figure}

The geometry of the test-case is illustrated in Fig. \ref{schem}. Photographs and renderings of the fabricated sample and holder are shown in Fig. 7. The hologram was a cylinder of radius of 2.54 cm and height of 3 cm (Fig.\ 6(a-b) and 7(a)) discretised on a 200 $\mu m$ grid. This was embedded in a separate 3.14x3.14x3 cm holder fabricated from veroClear (Fig.\ 6(a-b) and 7(b)). The input fields were each generated by a 2.5 cm, 2.25 MHz, PZT piston transducer (Olympus, Japan). This was attached to one side of the veroClear holder 5 mm below the top of the cylinder. The target plane was located parallel with the top of the cylinder at a depth of 1 cm (Fig.\ 6(b)), the design frequency was 2 MHz. To generate 8 distinct input fields from the fixed transducer, the cylindrical hologram was rotated within the buffer by 45\degree{} increments thereby changing the incidence angle of the piston field relative to the hologram (Fig. 6(c)). 

\subsection{Hologram calculation}
As stated the target fields consisted of the numerals `1' to `8'. These were decomposed into a set of 992 target points. Precisely calculating the Green's functions for these target points would require running a large set of full-wave simulations. To avoid this the hologram medium/water interface was approximated as an infinite half space neglecting reverberations within the hologram and between the hologram and veroClear buffer. A single simulation was then used to evaluate the Greens functions for a point source above this half space. The domain for this simulation was a 512$\times$32$\times$240 grid with a spacing of 200 $\mu m$ with a PML size of 10 grid points in each dimension. The hologram was inserted as a half-space with a height of 3 cm and acoustic properties $(c_m,\alpha_m,\rho_m)$. The rest of the medium was assigned to water. The point source was inserted as a 2 MHz additive source 1 cm above the interface. The steady state field $G(d,l)$ as a function of depth $d$ below the interface and radial distance $l$ from the source was extracted from this simulation. The values for $G_n^m(\mathbf{r}_k)$ were then calculated from $G(d,l)$ via interpolation. 

To calculate the input fields $p_i^n(\mathbf{r}_k)$ the PZT transducer was modelled as a piston with a diameter of 2.54 cm and a uniform apodisation and phase. The angular spectrum method (based on the implementation by Zeng and McGough\cite{Zeng2008}) was used to propagate this field from 0-3 cm on planes spaced by 200 $\mu m$. The medium properties were assigned to the bulk properties $(c_m,\alpha_m,\rho_m)$ of the hologram. From this 3D volume the position of each hologram coordinate $\mathbf{r}_k$, for each input field, was evaluated from the known transducer position, $p_i^n(\mathbf{r}_k)$ was then calculated via interpolation. After evaluation of the input fields and Greens functions the optimisation was run to convergence as outlined in Sec.\ IV. 

\subsection{Hologram Contrast}
As stated in Sec.\ III, the materials available by default on the Stratasys J835 have insufficient contrast to realise a volume hologram with high diffraction efficiency. Therefore, 'mixed' materials were synthesised using mixtures of Agilus30 ($c$ = 2035 m$\cdot$s$^{-1}$) and veroClear (2475 m$\cdot$s$^{-1}$) on a 200 $\mu m$ length scale. The hologram was then fabricated from this `mixed' material and veroClear. The target sound speed contrast $\Delta c$ was 75 m$\cdot$s$^{-1}$. This was empirically determined by calculating several holograms for the chosen geometry then numerically evaluating their output fields using k-Wave. For contrasts higher than 75 m$\cdot$s$^{-1}$ it was found that multiple scattering caused the forward model to increasingly break down. This gave a target sound speed for the mixed material of 2325 m$\cdot$s$^{-1}$ resulting in a mixing ratio of 34$\%$ Agilus30 and 66$\%$ veroClear. The bulk acoustic properties ($c_m$,$\alpha_m$,$\rho_m$) used for the hologram design were calculated by averaging overall volume fraction of Agilus30 and veroClear assuming a 50/50 mix of the `mixed' material and veroClear. The optimisation was then run to convergence for these finalised parameters.

\subsection{Fabrication and Measurement}
The hologram was next converted into Standard Tessellation Language (STL) for printing along with a model of the the veroClear buffer (Fig.\ 7(a-c)). Prior to this a circular ring with radius 45 mm and 8 clearance holes spaced by 45\degree was added to the hologram and 2 clearance holes added to the buffer to allow them to be attached with the appropriate set of orientations. Both parts were then fabricated on a Stratasys J835. An additional part, designed to hold the transducer, was printed on an Ultimaker 3 from PLA. 

The output field was measured for each hologram position in a custom built test tank with a three-axis computer controlled positioning system using a 0.5 mm needle hydrophone (Precision Acoustics, Dorchester). The cylindrical hologram was inserted in each of the 8 pre-set positions and the transducer was attached to the veroClear buffer. The transducer was driven with a 10 Vpp 15 cycle 2 MHz tone burst generated from a signal generator (33522A, Agilent Technologies, Santa Clara, CA, USA). The field was recorded over a 30x30 mm area, centered on the middle of the cylinder, at a depth of approximately 1 cm from the output surface, with a step size of 0.25 mm. Signals were recorded using a M4i.4450-x8 digitiser (Spectrum, Germany) using 64 averages. The acoustic field at 2 MHz was extracted at each measurement position.

\begin{figure}
    \centering
    \includegraphics[width=\columnwidth]{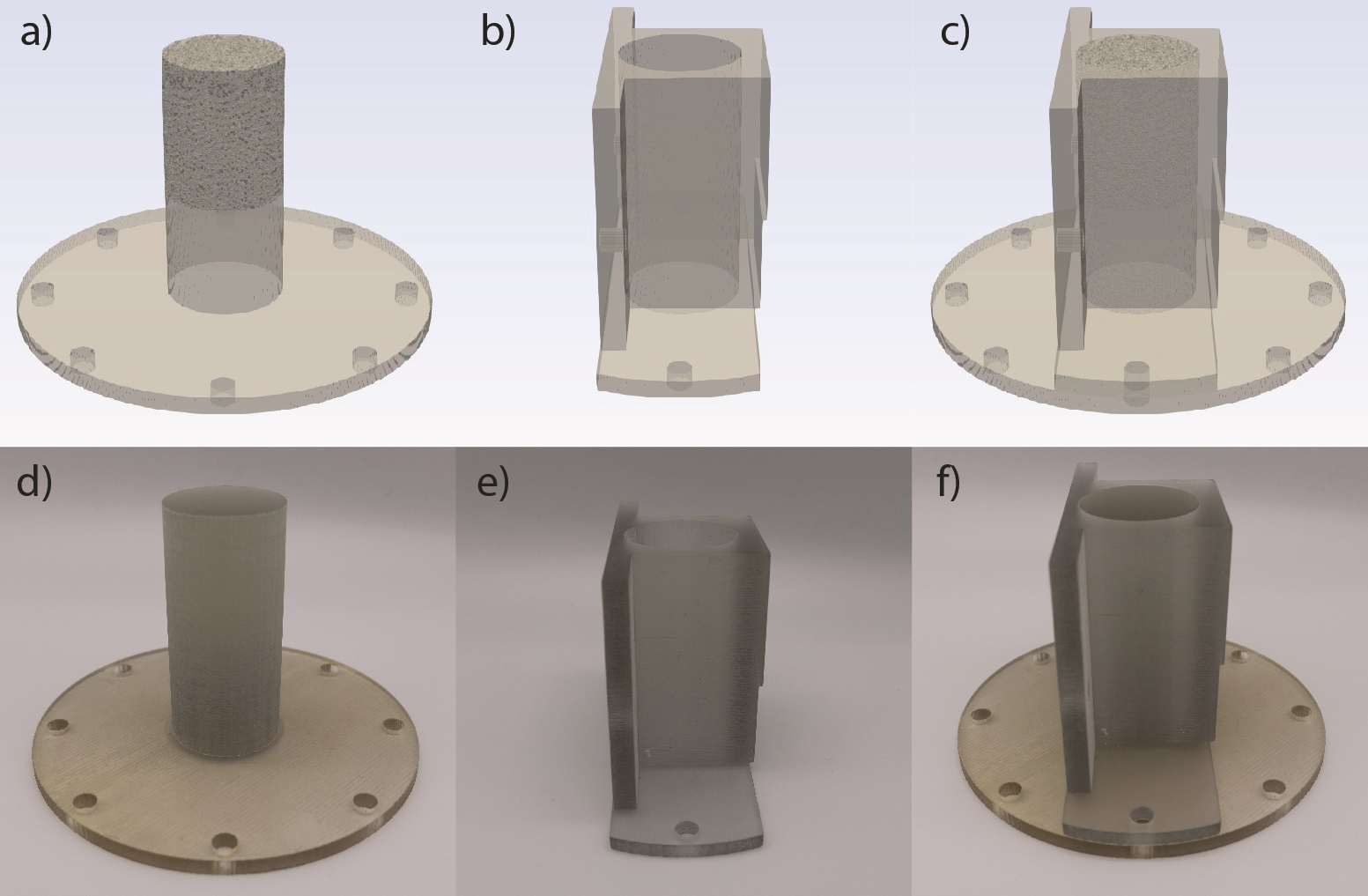}
    \caption{(a-c) Renderings of (a) Multiplexed volume hologram, (b) veroClear holder, (c) The two models attached to each other. (d-f) Photographs of fabricated models for (d) Volume hologram, (e) veroClear holder, (f) The two models attached to each other.}
    \label{fig_traces}
\end{figure}

Each field scan was compared against a full-wave simulation carried out using the k-Wave toolbox avoiding the simplifying assumption used to design the volume hologram. These simulations used a 256x256x240 grid and a step size of 200 $\mu m$. The time-step was 10 ns and the simulation length was 100 $\mu$s. Both the hologram and veroClear buffer were exactly inserted into the domain (i.e., each voxel was assigned Agilus30 or veroClear rather than approximating the medium with average bulk properties) and the rest of the medium was set to water. The PZT transducer was modelled, in each case, as a 2.54 cm additive piston source with uniform phase and apodisation driven at 2 MHz. The time-varying field was recorded over the target plane and the steady state field at 2 MHz extracted from each simulation. 

\begin{figure*}
    \centering
    \includegraphics[width=2\columnwidth]{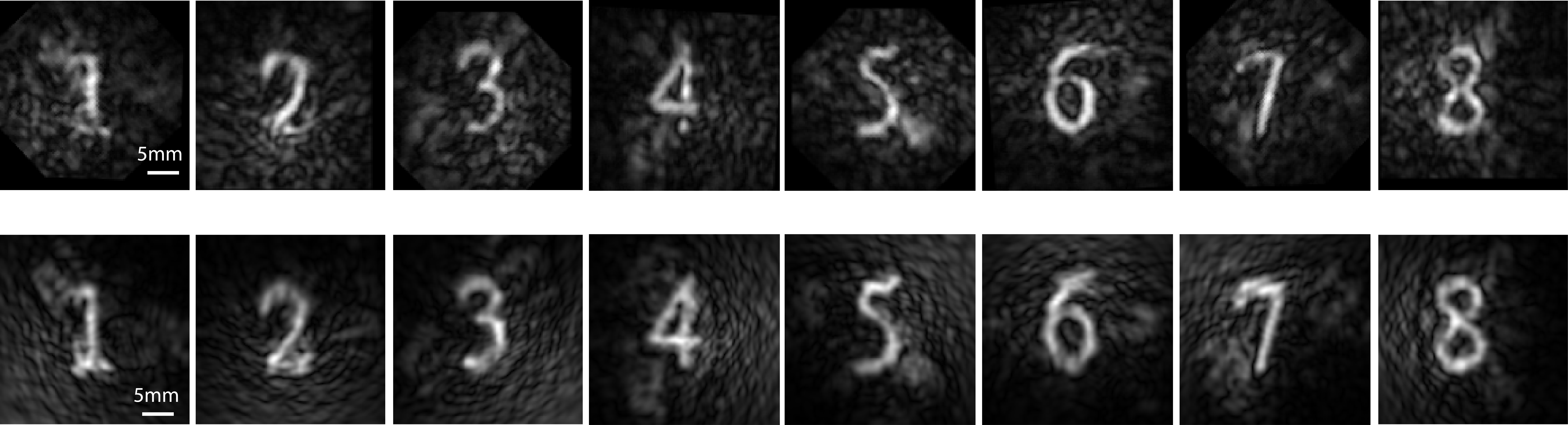}
    \caption{Experimental and simulation data for acoustic fields generated by binary volume acoustic hologram. Top row: Experimental measurements for each of the 8 fields at depth of 1 cm from output surface of hologram. Bottom row: Simulated results for each of the 8 fields at depth of 1 cm from output surface of hologram. Both simulated and experimental results generate the desired fields validating the generation of acoustic analogues of volumetric holograms}
    \label{fig_traces}
\end{figure*}

\subsection{Measurement Results}
The experimental (Top row) and simulation results (Bottom row) for each of the 8 target output fields can be seen in Fig.\ 8. The desired numeral is clearly generated in each experimental data-set confirming both that multi-polymer 3D printing can be used to realise acoustic analogues of volume holograms and that the assumptions of the design method are valid. The experimental measurements show good agreement with full-wave simulation demonstrating that k-Wave can be used to accurately predict the behaviour of these volume holograms. There are slightly greater distortions in some of the experimental measurements (e.g, the `1') compared to the simulated results. This could be due to several factors, for example, inaccuracies in the 3D printing process, inaccurate modelling of the transducer boundary conditions, or errors in material properties. 

\section{Summary and discussion}
This work has demonstrated that multiple-polymer 3D printing can be used to realise acoustic analogues to volume holograms and that these enable multiplexing of distinct output fields onto different incident field directions. A one-step algorithm for calculating these volume holograms in continuous media has been introduced along with a greedy optimisation approach capable of designing binary acoustic holograms in absorbing media. Both algorithms rely on a single scattering approximation, however, it has been shown experimentally that this holds for low-contrast. 

The encoding of 8 distinct patterns in a single lens is an improvement compared to previous demonstrations using single thin holographic lenses\cite{Brown2017}. As such, this could prove useful for applications already utilising acoustic holograms such as ultrasound neuromodulation\cite{Maimbourg2020} or blood-brain barrier opening\cite{Jimenez-Gambin2022}. The principle limitation of the approach in this work is the set of approximations employed in the forward model. Future work utilising approaches analogous to full-wave inversion could bypass this limitation allowing for higher-fidelity patterns, phase and amplitude control, and a greater ability to multiplex distinct distributions into a fixed lens. The high attenuation of 3D printable photopolymers also limits the efficiency with which energy can be coupled into the desired fields. In the future alternative fabrication methods employing materials with lower attenuation could bypass this drawback.

\begin{acknowledgments}
This work was supported by the Engineering and Physical Sciences Research Council, UK
\end{acknowledgments}



\bibliography{Manuscript.bib}

\begin{thebibliography}{35}%
\makeatletter
\providecommand \@ifxundefined [1]{%
 \@ifx{#1\undefined}
}%
\providecommand \@ifnum [1]{%
 \ifnum #1\expandafter \@firstoftwo
 \else \expandafter \@secondoftwo
 \fi
}%
\providecommand \@ifx [1]{%
 \ifx #1\expandafter \@firstoftwo
 \else \expandafter \@secondoftwo
 \fi
}%
\providecommand \natexlab [1]{#1}%
\providecommand \enquote  [1]{``#1''}%
\providecommand \bibnamefont  [1]{#1}%
\providecommand \bibfnamefont [1]{#1}%
\providecommand \citenamefont [1]{#1}%
\providecommand \href@noop [0]{\@secondoftwo}%
\providecommand \href [0]{\begingroup \@sanitize@url \@href}%
\providecommand \@href[1]{\@@startlink{#1}\@@href}%
\providecommand \@@href[1]{\endgroup#1\@@endlink}%
\providecommand \@sanitize@url [0]{\catcode `\\12\catcode `\$12\catcode
  `\&12\catcode `\#12\catcode `\^12\catcode `\_12\catcode `\%12\relax}%
\providecommand \@@startlink[1]{}%
\providecommand \@@endlink[0]{}%
\providecommand \url  [0]{\begingroup\@sanitize@url \@url }%
\providecommand \@url [1]{\endgroup\@href {#1}{\urlprefix }}%
\providecommand \urlprefix  [0]{URL }%
\providecommand \Eprint [0]{\href }%
\providecommand \doibase [0]{https://doi.org/}%
\providecommand \selectlanguage [0]{\@gobble}%
\providecommand \bibinfo  [0]{\@secondoftwo}%
\providecommand \bibfield  [0]{\@secondoftwo}%
\providecommand \translation [1]{[#1]}%
\providecommand \BibitemOpen [0]{}%
\providecommand \bibitemStop [0]{}%
\providecommand \bibitemNoStop [0]{.\EOS\space}%
\providecommand \EOS [0]{\spacefactor3000\relax}%
\providecommand \BibitemShut  [1]{\csname bibitem#1\endcsname}%
\let\auto@bib@innerbib\@empty
\bibitem [{\citenamefont {Melde}\ \emph {et~al.}(2018)\citenamefont {Melde},
  \citenamefont {Choi}, \citenamefont {Wu}, \citenamefont {Palagi},
  \citenamefont {Qiu},\ and\ \citenamefont {Fischer}}]{Melde2018}%
  \BibitemOpen
  \bibfield  {author} {\bibinfo {author} {\bibfnamefont {K.}~\bibnamefont
  {Melde}}, \bibinfo {author} {\bibfnamefont {E.}~\bibnamefont {Choi}},
  \bibinfo {author} {\bibfnamefont {Z.}~\bibnamefont {Wu}}, \bibinfo {author}
  {\bibfnamefont {S.}~\bibnamefont {Palagi}}, \bibinfo {author} {\bibfnamefont
  {T.}~\bibnamefont {Qiu}},\ and\ \bibinfo {author} {\bibfnamefont
  {P.}~\bibnamefont {Fischer}},\ }\bibfield  {title} {\bibinfo {title}
  {{Acoustic Fabrication via the Assembly and Fusion of Particles}},\
  }\bibfield  {journal} {\bibinfo  {journal} {Advanced Materials}\ }\textbf
  {\bibinfo {volume} {30}},\ \href {https://doi.org/10.1002/adma.201704507}
  {10.1002/adma.201704507} (\bibinfo {year} {2018})\BibitemShut {NoStop}%
\bibitem [{\citenamefont {Baudoin}\ \emph {et~al.}(2020)\citenamefont
  {Baudoin}, \citenamefont {Thomas}, \citenamefont {Sahely}, \citenamefont
  {Gerbedoen}, \citenamefont {Gong}, \citenamefont {Sivery}, \citenamefont
  {Matar}, \citenamefont {Smagin}, \citenamefont {Favreau},\ and\ \citenamefont
  {Vlandas}}]{Baudoin2020}%
  \BibitemOpen
  \bibfield  {author} {\bibinfo {author} {\bibfnamefont {M.}~\bibnamefont
  {Baudoin}}, \bibinfo {author} {\bibfnamefont {J.~L.}\ \bibnamefont {Thomas}},
  \bibinfo {author} {\bibfnamefont {R.~A.}\ \bibnamefont {Sahely}}, \bibinfo
  {author} {\bibfnamefont {J.~C.}\ \bibnamefont {Gerbedoen}}, \bibinfo {author}
  {\bibfnamefont {Z.}~\bibnamefont {Gong}}, \bibinfo {author} {\bibfnamefont
  {A.}~\bibnamefont {Sivery}}, \bibinfo {author} {\bibfnamefont {O.~B.}\
  \bibnamefont {Matar}}, \bibinfo {author} {\bibfnamefont {N.}~\bibnamefont
  {Smagin}}, \bibinfo {author} {\bibfnamefont {P.}~\bibnamefont {Favreau}},\
  and\ \bibinfo {author} {\bibfnamefont {A.}~\bibnamefont {Vlandas}},\
  }\bibfield  {title} {\bibinfo {title} {{Spatially selective manipulation of
  cells with single-beam acoustical tweezers}},\ }\href
  {https://doi.org/10.1038/s41467-020-18000-y} {\bibfield  {journal} {\bibinfo
  {journal} {Nature Communications}\ }\textbf {\bibinfo {volume} {11}},\
  \bibinfo {pages} {1} (\bibinfo {year} {2020})}\BibitemShut {NoStop}%
\bibitem [{\citenamefont {Kruizinga}\ \emph {et~al.}(2017)\citenamefont
  {Kruizinga}, \citenamefont {{Pim van der Meulen}}, \citenamefont {Fedjajevs},
  \citenamefont {Mastik}, \citenamefont {Springeling}, \citenamefont {{Nico de
  Jong}}, \citenamefont {Bosch},\ and\ \citenamefont {Leus}}]{Kruizinga2017}%
  \BibitemOpen
  \bibfield  {author} {\bibinfo {author} {\bibfnamefont {P.}~\bibnamefont
  {Kruizinga}}, \bibinfo {author} {\bibnamefont {{Pim van der Meulen}}},
  \bibinfo {author} {\bibfnamefont {A.}~\bibnamefont {Fedjajevs}}, \bibinfo
  {author} {\bibfnamefont {F.}~\bibnamefont {Mastik}}, \bibinfo {author}
  {\bibfnamefont {G.}~\bibnamefont {Springeling}}, \bibinfo {author}
  {\bibnamefont {{Nico de Jong}}}, \bibinfo {author} {\bibfnamefont {J.~G.}\
  \bibnamefont {Bosch}},\ and\ \bibinfo {author} {\bibfnamefont
  {G.}~\bibnamefont {Leus}},\ }\bibfield  {title} {\bibinfo {title}
  {{Compressive 3D ultrasound imaging using a single sensor}},\ }\bibfield
  {journal} {\bibinfo  {journal} {Science Advances}\ }\textbf {\bibinfo
  {volume} {3}},\ \href {https://doi.org/10.1126/sciadv.1701423}
  {10.1126/sciadv.1701423} (\bibinfo {year} {2017})\BibitemShut {NoStop}%
\bibitem [{\citenamefont {Maimbourg}\ \emph {et~al.}(2018)\citenamefont
  {Maimbourg}, \citenamefont {Houdouin}, \citenamefont {Deffieux},
  \citenamefont {Tanter},\ and\ \citenamefont {Aubry}}]{Maimbourg2018}%
  \BibitemOpen
  \bibfield  {author} {\bibinfo {author} {\bibfnamefont {G.}~\bibnamefont
  {Maimbourg}}, \bibinfo {author} {\bibfnamefont {A.}~\bibnamefont {Houdouin}},
  \bibinfo {author} {\bibfnamefont {T.}~\bibnamefont {Deffieux}}, \bibinfo
  {author} {\bibfnamefont {M.}~\bibnamefont {Tanter}},\ and\ \bibinfo {author}
  {\bibfnamefont {J.~F.}\ \bibnamefont {Aubry}},\ }\bibfield  {title} {\bibinfo
  {title} {{3D-printed adaptive acoustic lens as a disruptive technology for
  transcranial ultrasound therapy using single-element transducers}},\
  }\bibfield  {journal} {\bibinfo  {journal} {Physics in Medicine and Biology}\
  }\textbf {\bibinfo {volume} {63}},\ \href
  {https://doi.org/10.1088/1361-6560/aaa037} {10.1088/1361-6560/aaa037}
  (\bibinfo {year} {2018})\BibitemShut {NoStop}%
\bibitem [{\citenamefont {Kyriakou}\ \emph {et~al.}(2014)\citenamefont
  {Kyriakou}, \citenamefont {Neufeld}, \citenamefont {Werner}, \citenamefont
  {Paulides}, \citenamefont {Szekely},\ and\ \citenamefont
  {Kuster}}]{Kyriakou2014}%
  \BibitemOpen
  \bibfield  {author} {\bibinfo {author} {\bibfnamefont {A.}~\bibnamefont
  {Kyriakou}}, \bibinfo {author} {\bibfnamefont {E.}~\bibnamefont {Neufeld}},
  \bibinfo {author} {\bibfnamefont {B.}~\bibnamefont {Werner}}, \bibinfo
  {author} {\bibfnamefont {M.~M.}\ \bibnamefont {Paulides}}, \bibinfo {author}
  {\bibfnamefont {G.}~\bibnamefont {Szekely}},\ and\ \bibinfo {author}
  {\bibfnamefont {N.}~\bibnamefont {Kuster}},\ }\bibfield  {title} {\bibinfo
  {title} {{A review of numerical and experimental compensation techniques for
  skull-induced phase aberrations in transcranial focused ultrasound}},\ }\href
  {https://doi.org/10.3109/02656736.2013.861519} {\bibfield  {journal}
  {\bibinfo  {journal} {International Journal of Hyperthermia}\ }\textbf
  {\bibinfo {volume} {30}},\ \bibinfo {pages} {36} (\bibinfo {year}
  {2014})}\BibitemShut {NoStop}%
\bibitem [{\citenamefont {Memoli}\ \emph {et~al.}(2017)\citenamefont {Memoli},
  \citenamefont {Caleap}, \citenamefont {Asakawa}, \citenamefont {Sahoo},
  \citenamefont {Drinkwater},\ and\ \citenamefont {Subramanian}}]{Memoli2017}%
  \BibitemOpen
  \bibfield  {author} {\bibinfo {author} {\bibfnamefont {G.}~\bibnamefont
  {Memoli}}, \bibinfo {author} {\bibfnamefont {M.}~\bibnamefont {Caleap}},
  \bibinfo {author} {\bibfnamefont {M.}~\bibnamefont {Asakawa}}, \bibinfo
  {author} {\bibfnamefont {D.~R.}\ \bibnamefont {Sahoo}}, \bibinfo {author}
  {\bibfnamefont {B.~W.}\ \bibnamefont {Drinkwater}},\ and\ \bibinfo {author}
  {\bibfnamefont {S.}~\bibnamefont {Subramanian}},\ }\bibfield  {title}
  {\bibinfo {title} {{Metamaterial bricks and quantization of meta-surfaces}},\
  }\href {https://doi.org/10.1038/ncomms14608} {\bibfield  {journal} {\bibinfo
  {journal} {Nature Communications}\ }\textbf {\bibinfo {volume} {8}},\
  \bibinfo {pages} {1} (\bibinfo {year} {2017})}\BibitemShut {NoStop}%
\bibitem [{\citenamefont {Jim{\'{e}}nez-Gamb{\'{i}}n}\ \emph
  {et~al.}(2019)\citenamefont {Jim{\'{e}}nez-Gamb{\'{i}}n}, \citenamefont
  {Jim{\'{e}}nez}, \citenamefont {Benlloch},\ and\ \citenamefont
  {Camarena}}]{Jimenez-Gambin2019}%
  \BibitemOpen
  \bibfield  {author} {\bibinfo {author} {\bibfnamefont {S.}~\bibnamefont
  {Jim{\'{e}}nez-Gamb{\'{i}}n}}, \bibinfo {author} {\bibfnamefont
  {N.}~\bibnamefont {Jim{\'{e}}nez}}, \bibinfo {author} {\bibfnamefont {J.~M.}\
  \bibnamefont {Benlloch}},\ and\ \bibinfo {author} {\bibfnamefont
  {F.}~\bibnamefont {Camarena}},\ }\bibfield  {title} {\bibinfo {title}
  {{Holograms to Focus Arbitrary Ultrasonic Fields through the Skull}},\ }\href
  {https://doi.org/10.1103/PhysRevApplied.12.014016} {\bibfield  {journal}
  {\bibinfo  {journal} {Physical Review Applied}\ }\textbf {\bibinfo {volume}
  {12}},\ \bibinfo {pages} {1} (\bibinfo {year} {2019})},\ \Eprint
  {https://arxiv.org/abs/1902.06716} {arXiv:1902.06716} \BibitemShut {NoStop}%
\bibitem [{\citenamefont {Lalonde}\ \emph {et~al.}(1991)\citenamefont
  {Lalonde}, \citenamefont {Worthington},\ and\ \citenamefont
  {Hunt}}]{Lalonde1991}%
  \BibitemOpen
  \bibfield  {author} {\bibinfo {author} {\bibfnamefont {R.~J.}\ \bibnamefont
  {Lalonde}}, \bibinfo {author} {\bibfnamefont {A.}~\bibnamefont
  {Worthington}},\ and\ \bibinfo {author} {\bibfnamefont {J.~W.}\ \bibnamefont
  {Hunt}},\ }\bibfield  {title} {\bibinfo {title} {{Field conjugate acoustic
  lenses for ultrasound hyperthermia}},\ }\href
  {https://doi.org/10.1109/ULTSYM.1991.234045} {\bibfield  {journal} {\bibinfo
  {journal} {Proceedings - IEEE Ultrasonics Symposium}\ }\textbf {\bibinfo
  {volume} {40}},\ \bibinfo {pages} {1339} (\bibinfo {year}
  {1991})}\BibitemShut {NoStop}%
\bibitem [{\citenamefont {Memoli}\ \emph {et~al.}(2019)\citenamefont {Memoli},
  \citenamefont {Chisari}, \citenamefont {Eccles}, \citenamefont {Caleap},
  \citenamefont {Drinkwater},\ and\ \citenamefont {Subramanian}}]{Memoli2019}%
  \BibitemOpen
  \bibfield  {author} {\bibinfo {author} {\bibfnamefont {G.}~\bibnamefont
  {Memoli}}, \bibinfo {author} {\bibfnamefont {L.}~\bibnamefont {Chisari}},
  \bibinfo {author} {\bibfnamefont {J.~P.}\ \bibnamefont {Eccles}}, \bibinfo
  {author} {\bibfnamefont {M.}~\bibnamefont {Caleap}}, \bibinfo {author}
  {\bibfnamefont {B.~W.}\ \bibnamefont {Drinkwater}},\ and\ \bibinfo {author}
  {\bibfnamefont {S.}~\bibnamefont {Subramanian}},\ }\bibfield  {title}
  {\bibinfo {title} {{Vari-Sound: A varifocal lens for sound}},\ }\href
  {https://doi.org/10.1145/3290605.3300713} {\bibfield  {journal} {\bibinfo
  {journal} {Conference on Human Factors in Computing Systems - Proceedings}\
  ,\ \bibinfo {pages} {1}} (\bibinfo {year} {2019})}\BibitemShut {NoStop}%
\bibitem [{\citenamefont {Cox}\ \emph {et~al.}(2019)\citenamefont {Cox},
  \citenamefont {Melde}, \citenamefont {Croxford}, \citenamefont {Fischer},\
  and\ \citenamefont {Drinkwater}}]{Cox2019}%
  \BibitemOpen
  \bibfield  {author} {\bibinfo {author} {\bibfnamefont {L.}~\bibnamefont
  {Cox}}, \bibinfo {author} {\bibfnamefont {K.}~\bibnamefont {Melde}}, \bibinfo
  {author} {\bibfnamefont {A.}~\bibnamefont {Croxford}}, \bibinfo {author}
  {\bibfnamefont {P.}~\bibnamefont {Fischer}},\ and\ \bibinfo {author}
  {\bibfnamefont {B.~W.}\ \bibnamefont {Drinkwater}},\ }\bibfield  {title}
  {\bibinfo {title} {{Acoustic Hologram Enhanced Phased Arrays for Ultrasonic
  Particle Manipulation}},\ }\href
  {https://doi.org/10.1103/PhysRevApplied.12.064055} {\bibfield  {journal}
  {\bibinfo  {journal} {Physical Review Applied}\ }\textbf {\bibinfo {volume}
  {12}},\ \bibinfo {pages} {1} (\bibinfo {year} {2019})}\BibitemShut {NoStop}%
\bibitem [{\citenamefont {Lalonde}\ and\ \citenamefont
  {Hunt}(1995)}]{Lalonde1995}%
  \BibitemOpen
  \bibfield  {author} {\bibinfo {author} {\bibfnamefont {R.}~\bibnamefont
  {Lalonde}}\ and\ \bibinfo {author} {\bibfnamefont {J.~W.}\ \bibnamefont
  {Hunt}},\ }\bibfield  {title} {\bibinfo {title} {{Variable Frequency Field
  Conjugate Lenses for Ultrasound Hyperthermia}},\ }\href
  {https://doi.org/10.1109/58.464838} {\bibfield  {journal} {\bibinfo
  {journal} {IEEE Transactions on Ultrasonics, Ferroelectrics, and Frequency
  Control}\ }\textbf {\bibinfo {volume} {42}},\ \bibinfo {pages} {825}
  (\bibinfo {year} {1995})}\BibitemShut {NoStop}%
\bibitem [{\citenamefont {Brown}\ \emph {et~al.}(2020)\citenamefont {Brown},
  \citenamefont {Cox},\ and\ \citenamefont {Treeby}}]{Brown2020}%
  \BibitemOpen
  \bibfield  {author} {\bibinfo {author} {\bibfnamefont {M.~D.}\ \bibnamefont
  {Brown}}, \bibinfo {author} {\bibfnamefont {B.~T.}\ \bibnamefont {Cox}},\
  and\ \bibinfo {author} {\bibfnamefont {B.~E.}\ \bibnamefont {Treeby}},\
  }\bibfield  {title} {\bibinfo {title} {{Stackable acoustic holograms}},\
  }\bibfield  {journal} {\bibinfo  {journal} {Applied Physics Letters}\
  }\textbf {\bibinfo {volume} {116}},\ \href
  {https://doi.org/10.1063/5.0009829} {10.1063/5.0009829} (\bibinfo {year}
  {2020})\BibitemShut {NoStop}%
\bibitem [{\citenamefont {Brown}\ \emph {et~al.}(2017)\citenamefont {Brown},
  \citenamefont {Cox},\ and\ \citenamefont {Treeby}}]{Brown2017}%
  \BibitemOpen
  \bibfield  {author} {\bibinfo {author} {\bibfnamefont {M.~D.}\ \bibnamefont
  {Brown}}, \bibinfo {author} {\bibfnamefont {B.~T.}\ \bibnamefont {Cox}},\
  and\ \bibinfo {author} {\bibfnamefont {B.~E.}\ \bibnamefont {Treeby}},\
  }\bibfield  {title} {\bibinfo {title} {{Design of multi-frequency acoustic
  kinoforms}},\ }\bibfield  {journal} {\bibinfo  {journal} {Applied Physics
  Letters}\ }\textbf {\bibinfo {volume} {111}},\ \href
  {https://doi.org/10.1063/1.5004040} {10.1063/1.5004040} (\bibinfo {year}
  {2017})\BibitemShut {NoStop}%
\bibitem [{\citenamefont {Ma}\ \emph {et~al.}(2020)\citenamefont {Ma},
  \citenamefont {Melde}, \citenamefont {Athanassiadis}, \citenamefont {Schau},
  \citenamefont {Richter}, \citenamefont {Qiu},\ and\ \citenamefont
  {Fischer}}]{Ma2020}%
  \BibitemOpen
  \bibfield  {author} {\bibinfo {author} {\bibfnamefont {Z.}~\bibnamefont
  {Ma}}, \bibinfo {author} {\bibfnamefont {K.}~\bibnamefont {Melde}}, \bibinfo
  {author} {\bibfnamefont {A.~G.}\ \bibnamefont {Athanassiadis}}, \bibinfo
  {author} {\bibfnamefont {M.}~\bibnamefont {Schau}}, \bibinfo {author}
  {\bibfnamefont {H.}~\bibnamefont {Richter}}, \bibinfo {author} {\bibfnamefont
  {T.}~\bibnamefont {Qiu}},\ and\ \bibinfo {author} {\bibfnamefont
  {P.}~\bibnamefont {Fischer}},\ }\bibfield  {title} {\bibinfo {title}
  {{Spatial ultrasound modulation by digitally controlling microbubble
  arrays}},\ }\href {https://doi.org/10.1038/s41467-020-18347-2} {\bibfield
  {journal} {\bibinfo  {journal} {Nature Communications}\ }\textbf {\bibinfo
  {volume} {11}},\ \bibinfo {pages} {1} (\bibinfo {year} {2020})}\BibitemShut
  {NoStop}%
\bibitem [{\citenamefont {Ma}\ \emph {et~al.}(2022)\citenamefont {Ma},
  \citenamefont {Joh}, \citenamefont {Fan},\ and\ \citenamefont
  {Fischer}}]{Ma2022}%
  \BibitemOpen
  \bibfield  {author} {\bibinfo {author} {\bibfnamefont {Z.}~\bibnamefont
  {Ma}}, \bibinfo {author} {\bibfnamefont {H.}~\bibnamefont {Joh}}, \bibinfo
  {author} {\bibfnamefont {D.~E.}\ \bibnamefont {Fan}},\ and\ \bibinfo {author}
  {\bibfnamefont {P.}~\bibnamefont {Fischer}},\ }\bibfield  {title} {\bibinfo
  {title} {{Dynamic Ultrasound Projector Controlled by Light}},\ }\href
  {https://doi.org/10.1002/advs.202104401} {\bibfield  {journal} {\bibinfo
  {journal} {Advanced Science}\ }\textbf {\bibinfo {volume} {9}},\ \bibinfo
  {pages} {1} (\bibinfo {year} {2022})}\BibitemShut {NoStop}%
\bibitem [{\citenamefont {Moreno}\ \emph {et~al.}(1997)\citenamefont {Moreno},
  \citenamefont {Rom{\'{a}}n},\ and\ \citenamefont {Salgueiro}}]{Moreno1997}%
  \BibitemOpen
  \bibfield  {author} {\bibinfo {author} {\bibfnamefont {V.}~\bibnamefont
  {Moreno}}, \bibinfo {author} {\bibfnamefont {J.~F.}\ \bibnamefont
  {Rom{\'{a}}n}},\ and\ \bibinfo {author} {\bibfnamefont {J.~R.}\ \bibnamefont
  {Salgueiro}},\ }\bibfield  {title} {\bibinfo {title} {{High efficiency
  diffractive lenses: Deduction of kinoform profile}},\ }\href
  {https://doi.org/10.1119/1.18587} {\bibfield  {journal} {\bibinfo  {journal}
  {American Journal of Physics}\ }\textbf {\bibinfo {volume} {65}},\ \bibinfo
  {pages} {556} (\bibinfo {year} {1997})}\BibitemShut {NoStop}%
\bibitem [{\citenamefont {Russell}(1981)}]{Russell1981}%
  \BibitemOpen
  \bibfield  {author} {\bibinfo {author} {\bibfnamefont {P.~S.~J.}\
  \bibnamefont {Russell}},\ }\bibfield  {title} {\bibinfo {title} {{Optical
  volume holography}},\ }\href {https://doi.org/10.1016/0370-1573(81)90196-4}
  {\bibfield  {journal} {\bibinfo  {journal} {Physics Reports}\ }\textbf
  {\bibinfo {volume} {71}},\ \bibinfo {pages} {209} (\bibinfo {year}
  {1981})}\BibitemShut {NoStop}%
\bibitem [{\citenamefont {van Heerden}(1963)}]{VanHeerden1963}%
  \BibitemOpen
  \bibfield  {author} {\bibinfo {author} {\bibfnamefont {P.~J.}\ \bibnamefont
  {van Heerden}},\ }\bibfield  {title} {\bibinfo {title} {{Theory of Optical
  Information Storage in Solids}},\ }\href
  {https://doi.org/10.1364/ao.2.000764} {\bibfield  {journal} {\bibinfo
  {journal} {Applied Optics}\ }\textbf {\bibinfo {volume} {2}},\ \bibinfo
  {pages} {764} (\bibinfo {year} {1963})}\BibitemShut {NoStop}%
\bibitem [{\citenamefont {P{\'{e}}rez-Liva}\ \emph {et~al.}(2017)\citenamefont
  {P{\'{e}}rez-Liva}, \citenamefont {Herraiz}, \citenamefont {Ud{\'{i}}as},
  \citenamefont {Miller}, \citenamefont {Cox},\ and\ \citenamefont
  {Treeby}}]{Perez-Liva2017}%
  \BibitemOpen
  \bibfield  {author} {\bibinfo {author} {\bibfnamefont {M.}~\bibnamefont
  {P{\'{e}}rez-Liva}}, \bibinfo {author} {\bibfnamefont {J.~L.}\ \bibnamefont
  {Herraiz}}, \bibinfo {author} {\bibfnamefont {J.~M.}\ \bibnamefont
  {Ud{\'{i}}as}}, \bibinfo {author} {\bibfnamefont {E.}~\bibnamefont {Miller}},
  \bibinfo {author} {\bibfnamefont {B.~T.}\ \bibnamefont {Cox}},\ and\ \bibinfo
  {author} {\bibfnamefont {B.~E.}\ \bibnamefont {Treeby}},\ }\bibfield  {title}
  {\bibinfo {title} {{Time domain reconstruction of sound speed and attenuation
  in ultrasound computed tomography using full wave inversion}},\ }\href
  {https://doi.org/10.1121/1.4976688} {\bibfield  {journal} {\bibinfo
  {journal} {The Journal of the Acoustical Society of America}\ }\textbf
  {\bibinfo {volume} {141}},\ \bibinfo {pages} {1595} (\bibinfo {year}
  {2017})}\BibitemShut {NoStop}%
\bibitem [{\citenamefont {Kogelnik}(1969)}]{Kogelnik1969}%
  \BibitemOpen
  \bibfield  {author} {\bibinfo {author} {\bibfnamefont {H.}~\bibnamefont
  {Kogelnik}},\ }\bibfield  {title} {\bibinfo {title} {{Coupled wave theory for
  thick hologram gratings}},\ }\href@noop {} {\bibfield  {journal} {\bibinfo
  {journal} {Bell System Technical Journal}\ }\textbf {\bibinfo {volume} {48}}
  (\bibinfo {year} {1969})}\BibitemShut {NoStop}%
\bibitem [{\citenamefont
  {Brotherton-Ratcliffe}(2013)}]{BrothertonRatcliffei2013}%
  \BibitemOpen
  \bibfield  {author} {\bibinfo {author} {\bibfnamefont {D.}~\bibnamefont
  {Brotherton-Ratcliffe}},\ }\bibfield  {title} {\bibinfo {title}
  {{Understanding Diffraction in Volume Gratings and Holograms}},\ }in\ \href
  {http://dx.doi.org/10.1039/C7RA00172J%0Ahttps://www.intechopen.com/books/advanced-biometric-technologies/liveness-detection-in-biometrics%0Ahttp://dx.doi.org/10.1016/j.colsurfa.2011.12.014}
  {\emph {\bibinfo {booktitle} {Holography: Basic Principles and Contemporary
  Applications}}}\ (\bibinfo {year} {2013})\ Chap.~\bibinfo {chapter} {1}, p.\
  \bibinfo {pages} {392}\BibitemShut {NoStop}%
\bibitem [{\citenamefont {Moharam}\ and\ \citenamefont
  {Gaylord}(1981)}]{Moharam1981}%
  \BibitemOpen
  \bibfield  {author} {\bibinfo {author} {\bibfnamefont {M.~G.}\ \bibnamefont
  {Moharam}}\ and\ \bibinfo {author} {\bibfnamefont {T.~K.}\ \bibnamefont
  {Gaylord}},\ }\bibfield  {title} {\bibinfo {title} {{Rigorous coupled-wave
  analysis of planar-grating diffraction}},\ }\href@noop {} {\bibfield
  {journal} {\bibinfo  {journal} {Journal of the Optical Society of America}\
  }\textbf {\bibinfo {volume} {71}} (\bibinfo {year} {1981})}\BibitemShut
  {NoStop}%
\bibitem [{\citenamefont {Treeby}\ and\ \citenamefont
  {Cox}(2010)}]{Treeby2010}%
  \BibitemOpen
  \bibfield  {author} {\bibinfo {author} {\bibfnamefont {B.~E.}\ \bibnamefont
  {Treeby}}\ and\ \bibinfo {author} {\bibfnamefont {B.~T.}\ \bibnamefont
  {Cox}},\ }\bibfield  {title} {\bibinfo {title} {{k-Wave: MATLAB toolbox for
  the simulation and reconstruction of photoacoustic wave fields}},\ }\href
  {https://doi.org/10.1117/1.3360308} {\bibfield  {journal} {\bibinfo
  {journal} {Journal of Biomedical Optics}\ }\textbf {\bibinfo {volume} {15}},\
  \bibinfo {pages} {021314} (\bibinfo {year} {2010})}\BibitemShut {NoStop}%
\bibitem [{\citenamefont {Heanue}\ \emph {et~al.}(1994)\citenamefont {Heanue},
  \citenamefont {Bashaw},\ and\ \citenamefont {Hesselink}}]{Heanue1994}%
  \BibitemOpen
  \bibfield  {author} {\bibinfo {author} {\bibfnamefont {J.~F.}\ \bibnamefont
  {Heanue}}, \bibinfo {author} {\bibfnamefont {M.~C.}\ \bibnamefont {Bashaw}},\
  and\ \bibinfo {author} {\bibfnamefont {L.}~\bibnamefont {Hesselink}},\
  }\bibfield  {title} {\bibinfo {title} {{Volume holographic storage and
  retrieval of digital data}},\ }\href
  {https://doi.org/10.1126/science.265.5173.749} {\bibfield  {journal}
  {\bibinfo  {journal} {Science}\ }\textbf {\bibinfo {volume} {265}},\ \bibinfo
  {pages} {749} (\bibinfo {year} {1994})}\BibitemShut {NoStop}%
\bibitem [{\citenamefont {Rosen}\ \emph {et~al.}(1993)\citenamefont {Rosen},
  \citenamefont {Segev},\ and\ \citenamefont {Yariv}}]{Rosen1993}%
  \BibitemOpen
  \bibfield  {author} {\bibinfo {author} {\bibfnamefont {J.}~\bibnamefont
  {Rosen}}, \bibinfo {author} {\bibfnamefont {M.}~\bibnamefont {Segev}},\ and\
  \bibinfo {author} {\bibfnamefont {A.}~\bibnamefont {Yariv}},\ }\bibfield
  {title} {\bibinfo {title} {{Wavelength-multiplexed computer-generated volume
  holography}},\ }\href {https://doi.org/10.1364/ol.18.000744} {\bibfield
  {journal} {\bibinfo  {journal} {Optics Letters}\ }\textbf {\bibinfo {volume}
  {18}},\ \bibinfo {pages} {744} (\bibinfo {year} {1993})}\BibitemShut
  {NoStop}%
\bibitem [{\citenamefont {Shishova}\ \emph {et~al.}(2020)\citenamefont
  {Shishova}, \citenamefont {Zherdev}, \citenamefont {Lushnikov},\ and\
  \citenamefont {Odinokov}}]{Shishova2020}%
  \BibitemOpen
  \bibfield  {author} {\bibinfo {author} {\bibfnamefont {M.}~\bibnamefont
  {Shishova}}, \bibinfo {author} {\bibfnamefont {A.}~\bibnamefont {Zherdev}},
  \bibinfo {author} {\bibfnamefont {D.}~\bibnamefont {Lushnikov}},\ and\
  \bibinfo {author} {\bibfnamefont {S.}~\bibnamefont {Odinokov}},\ }\bibfield
  {title} {\bibinfo {title} {{Recording of the multiplexed bragg diffraction
  gratings for waveguides using phase mask}},\ }\href
  {https://doi.org/10.3390/photonics7040097} {\bibfield  {journal} {\bibinfo
  {journal} {Photonics}\ }\textbf {\bibinfo {volume} {7}},\ \bibinfo {pages}
  {1} (\bibinfo {year} {2020})}\BibitemShut {NoStop}%
\bibitem [{\citenamefont {Burr}(1996)}]{Burr1996}%
  \BibitemOpen
  \bibfield  {author} {\bibinfo {author} {\bibfnamefont {G.~W.}\ \bibnamefont
  {Burr}},\ }\emph {\bibinfo {title} {{Volume holographic storage using the
  90$\degree$ geometry}}},\ \href@noop {} {Ph.D. thesis},\ \bibinfo  {school}
  {California Institute of Technology} (\bibinfo {year} {1996})\BibitemShut
  {NoStop}%
\bibitem [{\citenamefont {Gerke}\ and\ \citenamefont
  {Piestun}(2010)}]{Gerke2010}%
  \BibitemOpen
  \bibfield  {author} {\bibinfo {author} {\bibfnamefont {T.~D.}\ \bibnamefont
  {Gerke}}\ and\ \bibinfo {author} {\bibfnamefont {R.}~\bibnamefont
  {Piestun}},\ }\bibfield  {title} {\bibinfo {title} {Aperiodic volume
  optics},\ }\href {https://doi.org/10.1038/nphoton.2009.290} {\bibfield
  {journal} {\bibinfo  {journal} {Nature Photonics}\ }\textbf {\bibinfo
  {volume} {4}},\ \bibinfo {pages} {188} (\bibinfo {year} {2010})}\BibitemShut
  {NoStop}%
\bibitem [{\citenamefont {Vaezi}\ \emph {et~al.}(2013)\citenamefont {Vaezi},
  \citenamefont {Chianrabutra}, \citenamefont {Mellor},\ and\ \citenamefont
  {Yang}}]{Vaezi2013}%
  \BibitemOpen
  \bibfield  {author} {\bibinfo {author} {\bibfnamefont {M.}~\bibnamefont
  {Vaezi}}, \bibinfo {author} {\bibfnamefont {S.}~\bibnamefont {Chianrabutra}},
  \bibinfo {author} {\bibfnamefont {B.}~\bibnamefont {Mellor}},\ and\ \bibinfo
  {author} {\bibfnamefont {S.}~\bibnamefont {Yang}},\ }\bibfield  {title}
  {\bibinfo {title} {{Multiple material additive manufacturing - Part 1: A
  review}},\ }\href {https://doi.org/10.1080/17452759.2013.778175} {\bibfield
  {journal} {\bibinfo  {journal} {Virtual and Physical Prototyping}\ }\textbf
  {\bibinfo {volume} {8}},\ \bibinfo {pages} {19} (\bibinfo {year}
  {2013})}\BibitemShut {NoStop}%
\bibitem [{\citenamefont {Bakaric}\ \emph {et~al.}(2021)\citenamefont
  {Bakaric}, \citenamefont {Miloro}, \citenamefont {Javaherian}, \citenamefont
  {Cox}, \citenamefont {Treeby},\ and\ \citenamefont {Brown}}]{Bakaric2021}%
  \BibitemOpen
  \bibfield  {author} {\bibinfo {author} {\bibfnamefont {M.}~\bibnamefont
  {Bakaric}}, \bibinfo {author} {\bibfnamefont {P.}~\bibnamefont {Miloro}},
  \bibinfo {author} {\bibfnamefont {A.}~\bibnamefont {Javaherian}}, \bibinfo
  {author} {\bibfnamefont {B.~T.}\ \bibnamefont {Cox}}, \bibinfo {author}
  {\bibfnamefont {B.~E.}\ \bibnamefont {Treeby}},\ and\ \bibinfo {author}
  {\bibfnamefont {M.~D.}\ \bibnamefont {Brown}},\ }\bibfield  {title} {\bibinfo
  {title} {{Measurement of the ultrasound attenuation and dispersion in
  3D-printed photopolymer materials from 1 to 3.5 MHz}},\ }\href
  {https://doi.org/10.1121/10.0006668} {\bibfield  {journal} {\bibinfo
  {journal} {The Journal of the Acoustical Society of America}\ }\textbf
  {\bibinfo {volume} {150}},\ \bibinfo {pages} {2798} (\bibinfo {year}
  {2021})}\BibitemShut {NoStop}%
\bibitem [{\citenamefont {Tsang}\ \emph {et~al.}(2011)\citenamefont {Tsang},
  \citenamefont {Poon}, \citenamefont {Cheung},\ and\ \citenamefont
  {Liu}}]{Tsang2011}%
  \BibitemOpen
  \bibfield  {author} {\bibinfo {author} {\bibfnamefont {P.}~\bibnamefont
  {Tsang}}, \bibinfo {author} {\bibfnamefont {T.-C.}\ \bibnamefont {Poon}},
  \bibinfo {author} {\bibfnamefont {W.-K.}\ \bibnamefont {Cheung}},\ and\
  \bibinfo {author} {\bibfnamefont {J.-P.}\ \bibnamefont {Liu}},\ }\bibfield
  {title} {\bibinfo {title} {{Computer generation of binary Fresnel
  holography}},\ }\href {https://doi.org/10.1364/AO.50.000B88} {\bibfield
  {journal} {\bibinfo  {journal} {Applied Optics}\ }\textbf {\bibinfo {volume}
  {50}},\ \bibinfo {pages} {B88} (\bibinfo {year} {2011})}\BibitemShut
  {NoStop}%
\bibitem [{\citenamefont {Brown}\ \emph {et~al.}(2014)\citenamefont {Brown},
  \citenamefont {Allen}, \citenamefont {Cox},\ and\ \citenamefont
  {Treeby}}]{Brown2014}%
  \BibitemOpen
  \bibfield  {author} {\bibinfo {author} {\bibfnamefont {M.~D.}\ \bibnamefont
  {Brown}}, \bibinfo {author} {\bibfnamefont {T.~J.}\ \bibnamefont {Allen}},
  \bibinfo {author} {\bibfnamefont {B.~T.}\ \bibnamefont {Cox}},\ and\ \bibinfo
  {author} {\bibfnamefont {B.~E.}\ \bibnamefont {Treeby}},\ }\bibfield  {title}
  {\bibinfo {title} {{Control of optically generated ultrasound fields using
  binary amplitude holograms}},\ }\href
  {https://doi.org/10.1109/ULTSYM.2014.0254} {\bibfield  {journal} {\bibinfo
  {journal} {IEEE International Ultrasonics Symposium, IUS}\ }\textbf {\bibinfo
  {volume} {139}},\ \bibinfo {pages} {1037} (\bibinfo {year} {2014})},\ \Eprint
  {https://arxiv.org/abs/arXiv:1605.00133} {arXiv:arXiv:1605.00133}
  \BibitemShut {NoStop}%
\bibitem [{\citenamefont {Zeng}\ and\ \citenamefont
  {McGough}(2008)}]{Zeng2008}%
  \BibitemOpen
  \bibfield  {author} {\bibinfo {author} {\bibfnamefont {X.}~\bibnamefont
  {Zeng}}\ and\ \bibinfo {author} {\bibfnamefont {R.~J.}\ \bibnamefont
  {McGough}},\ }\bibfield  {title} {\bibinfo {title} {{Evaluation of the
  angular spectrum approach for simulations of near-field pressures}},\ }\href
  {https://doi.org/10.1016/j.neuron.2009.10.017.A} {\bibfield  {journal}
  {\bibinfo  {journal} {Journal Acoustical Society of America}\ }\textbf
  {\bibinfo {volume} {123}},\ \bibinfo {pages} {68} (\bibinfo {year} {2008})},\
  \Eprint {https://arxiv.org/abs/NIHMS150003} {arXiv:NIHMS150003} \BibitemShut
  {NoStop}%
\bibitem [{\citenamefont {Maimbourg}\ \emph {et~al.}(2020)\citenamefont
  {Maimbourg}, \citenamefont {Houdouin}, \citenamefont {Deffieux},
  \citenamefont {Tanter},\ and\ \citenamefont {Aubry}}]{Maimbourg2020}%
  \BibitemOpen
  \bibfield  {author} {\bibinfo {author} {\bibfnamefont {G.}~\bibnamefont
  {Maimbourg}}, \bibinfo {author} {\bibfnamefont {A.}~\bibnamefont {Houdouin}},
  \bibinfo {author} {\bibfnamefont {T.}~\bibnamefont {Deffieux}}, \bibinfo
  {author} {\bibfnamefont {M.}~\bibnamefont {Tanter}},\ and\ \bibinfo {author}
  {\bibfnamefont {J.~F.}\ \bibnamefont {Aubry}},\ }\bibfield  {title} {\bibinfo
  {title} {{Steering capabilities of an acoustic lens for transcranial therapy:
  Numerical and experimental studies}},\ }\href
  {https://doi.org/10.1109/TBME.2019.2907556} {\bibfield  {journal} {\bibinfo
  {journal} {IEEE Transactions on Biomedical Engineering}\ }\textbf {\bibinfo
  {volume} {67}},\ \bibinfo {pages} {27} (\bibinfo {year} {2020})}\BibitemShut
  {NoStop}%
\bibitem [{\citenamefont {Jimenez-Gambin}\ \emph {et~al.}(2022)\citenamefont
  {Jimenez-Gambin}, \citenamefont {Jimenez}, \citenamefont {Pouliopoulos},
  \citenamefont {Benlloch}, \citenamefont {Konofagou},\ and\ \citenamefont
  {Camarena}}]{Jimenez-Gambin2022}%
  \BibitemOpen
  \bibfield  {author} {\bibinfo {author} {\bibfnamefont {S.}~\bibnamefont
  {Jimenez-Gambin}}, \bibinfo {author} {\bibfnamefont {N.}~\bibnamefont
  {Jimenez}}, \bibinfo {author} {\bibfnamefont {A.}~\bibnamefont
  {Pouliopoulos}}, \bibinfo {author} {\bibfnamefont {J.~M.}\ \bibnamefont
  {Benlloch}}, \bibinfo {author} {\bibfnamefont {E.}~\bibnamefont
  {Konofagou}},\ and\ \bibinfo {author} {\bibfnamefont {F.}~\bibnamefont
  {Camarena}},\ }\bibfield  {title} {\bibinfo {title} {{Acoustic Holograms for
  Bilateral Blood-Brain Barrier Opening in a Mouse Model}},\ }\href
  {https://doi.org/10.1109/TBME.2021.3115553} {\bibfield  {journal} {\bibinfo
  {journal} {IEEE Transactions on Biomedical Engineering}\ }\textbf {\bibinfo
  {volume} {69}},\ \bibinfo {pages} {1359} (\bibinfo {year}
  {2022})}\BibitemShut {NoStop}%
\end{thebibliography}%

\end{document}